\documentclass[twocolumn]{aastex62}
\usepackage{enumitem}
\usepackage{graphicx,url}
\usepackage{color}
\usepackage{hyperref}
\usepackage{amsmath}
\usepackage{amssymb}
\usepackage{mathtools}
\usepackage[htt]{hyphenat}
\usepackage[utf8]{inputenc}
\usepackage{lineno}

\defcitealias{Zdziarski2021a}{Z21}
\defcitealias{Garcia2018a}{G18}

\def\j1550{XTE~J1550$-$564}

\def\u1630{4U~1630$-$47}
\def\j1752{XTE~J1752$-$223}

\def\rxte{{\it RXTE}}

\def\swift{{\it Swift}}
\def\nustar{{\it NuSTAR}}

\accepted{by ApJ, 2022 July 9}

\shorttitle{Reflection Spectroscopy of \j1752}
\shortauthors{Connors et al.}

\setlist[itemize]{leftmargin=*}

\begin{document}

\title{The long-stable hard state of \j1752 and the disk truncation dilemma}

\correspondingauthor{Riley~M.~T.~Connors}
\email{rconnors@caltech.edu}

\author{Riley~M.~T.~Connors}
\affiliation{Cahill Center for Astronomy and Astrophysics, California Institute of Technology, \\
  Pasadena, CA 91125, USA}
  
 \author{Javier~A.~Garc\'ia}
\affiliation{Cahill Center for Astronomy and Astrophysics, California Institute of Technology, \\
  Pasadena, CA 91125, USA}
  \affiliation{Dr Karl Remeis-Observatory and Erlangen Centre for Astroparticle Physics,\\
  Sternwartstr. 7, D-96049 Bamberg, Germany}
  
 \author{John~Tomsick}
\affiliation{Space Sciences Laboratory, University of California Berkeley,\\
7 Gauss Way, Berkeley, CA 94720-7450}

\author{Guglielmo~Mastroserio}
\affiliation{Cahill Center for Astronomy and Astrophysics, California Institute of Technology, \\
  Pasadena, CA 91125, USA}
  
\author{Victoria~Grinberg}
\affiliation{European Space Agency (ESA), European Space Research and Technology Centre (ESTEC), Keplerlaan 1, 2201 AZ Noordw?k, the Netherlands}

\author{James~F.~Steiner}
\affiliation{MIT Kavli Institute, 77 Massachusetts Avenue, 37-241, \\
Cambridge, MA 02139, USA}

\author{Jiachen~Jiang}
\affiliation{Institute of Astronomy, University of Cambridge, Madingley Road, Cambridge CB3 0HA, UK}

\author{Andrew~C.~Fabian}
\affiliation{Institute of Astronomy, University of Cambridge, Madingley Road, Cambridge CB3 0HA, UK}

\author{Michael~L.~Parker}
\affiliation{Institute of Astronomy, University of Cambridge, Madingley Road, Cambridge CB3 0HA, UK}

\author{Fiona~Harrison}
 \affiliation{Cahill Center for Astronomy and Astrophysics, California Institute of Technology, \\
  Pasadena, CA 91125, USA}
  
  \author{Jeremy~Hare}
\affiliation{NASA Postdoctoral Program Fellow, NASA Goddard Space Flight Center, Greenbelt, MD 20771, USA}

\author{Labani~Mallick}
\affiliation{Cahill Center for Astronomy and Astrophysics, California Institute of Technology, \\
  Pasadena, CA 91125, USA}
  
  \author{Hadar~Lazar}
\affiliation{Space Sciences Laboratory, University of California Berkeley,\\
7 Gauss Way, Berkeley, CA 94720-7450}
  
\begin{abstract}

The degree to which the thin accretion disks of black hole X-ray binaries are truncated during hard spectral states remains a contentious open question in black hole astrophysics. During its singular observed outburst in $2009\mbox{--}2010$, the black hole X-ray binary \j1752\ spent $\sim1$~month in a long-stable hard spectral state at a luminosity of $\sim0.02\mbox{--}0.1~L_{\rm Edd}$. It was observed with 56 \rxte\ pointings during this period, with simultaneous \swift-XRT daily coverage during the first 10 days of the \rxte\ observations. Whilst reflection modeling has been extensively explored in the analysis of these data, there is a disagreement surrounding the geometry of the accretion disk and corona implied by the reflection features. We re-examine the combined, high signal-to-noise, simultaneous \swift\ and \rxte\ observations, and perform extensive reflection modeling with the latest {\tt relxill} suite of reflection models, including newer high disk density models. We show that reflection modeling requires that the disk be within $\sim5~R_{\rm ISCO}$ during the hard spectral state, whilst weaker constraints from the thermal disk emission imply higher truncation ($R_{\rm in}=6\mbox{--}80~R_{\rm ISCO}$). We also explore more complex coronal continuum models, allowing for two Comptonization components instead of one, and show that the reflection features still require only a mildly truncated disk. Finally we present a full comparison of our results to previous constraints found from analyses of the same dataset. 

\end{abstract}


%
%
%
\section{Introduction}\label{sec:intro}

Black hole X-ray binaries (BHBs) are accreting compact objects that display a broad range of variable spectral and temporal variability on sub-second to year-long timescales. As such they are an ideal laboratory for the study of accretion physics and radiative processes in the strong gravity regime. For the same reasons they present an opportunity to understand their supermassive cousins in active galactic nuclei, objects that behave in similar ways (see, e.g., \citealt{Plotkin2012,Markoff2015,Connors2017}). 

A key unknown persists in studies of the evolution of BHB outbursts: during which stage of the outburst has the optically thick, geometrically thin accretion disk reached the innermost stable circular orbit (ISCO)? Two primary hypotheses remain: (i) as BHBs reach the bright hard state ($L_{\rm X} \ge 0.01~L_{\rm Edd}$) the disk has reached, or is close to, the ISCO (e.g., \citealt{Garcia2015}); (ii) the disks of BHBs do not extend to the ISCO until the source has evolved from the hard to the soft state, with predictions/measurements as extreme as $R_{\rm in}\sim100~R_{\rm ISCO}$ during bright hard states (e.g., \citealt{Zdziarski2020a}). Solving this disagreement is paramount to our fundamental understanding of BHs and accretion, since measurements of BH spin, $a_{\star}$, rely on the assumption that the reflecting material (i.e., the disk) is at---or close to---the ISCO \citep{Reynolds1997, Young1998,Reynolds2008,Fabian2014}.

The BHB \j1752 was discovered in October $2009$ by the {\it Rossi X-ray Timing Explorer} (\rxte) in its first and thus far only outburst, which ended in May $2010$ \citep{Markwardt2009b, Nakahira2009, Shaposhnikov2009, Shaposhnikov2010b,Shaposhnikov2010a}. Its spectral and timing properties, as well as the presence of persistent and ballistic radio jets, confirmed its nature as a BHB \citep{Brocksopp2010,Brocksopp2013,Yang2010,Yang2011}. \cite{Shaposhnikov2010a} estimated the BH mass and source distance via correlations of spectral and variability properties, yielding a BH mass of $\sim8\mbox{--}11~M_{\odot}$, and a distance of $\sim3.5$~kpc. Observations of the radio jet later revealed the jet velocity and inclination to the light of sight to be $\beta>0.66$ and $i<49^{\circ}$ respectively \citep{Miller-Jones2011}. 

Recent work on reflection modeling of \j1752 by \cite{Garcia2018a} (\citetalias{Garcia2018a} hereafter) showed that in the isolated and stable hard state observed by \rxte, the disk appears to be close to the ISCO. However, since the publication of \citetalias{Garcia2018a} there have been key developments in the breadth of reflection spectra models, in particular with regards to treatments of the accretion disk density, $n_{\rm e}$. Early implementations of the model {\tt relxill} \citep{Garcia2014,Dauser2014} assumed a canonical disk density of $n_{\rm e}=10^{15}~{\rm cm^{-3}}$, generally appropriate for disks around supermassive black holes. Calculations based on the standard Shakura-Sunyaev theory of accretion disk structure \citep{SS1973} predict disk densities around stellar mass black holes of $n>10^{20}~{\rm cm^{-3}}$, even for relatively high accretion rates ($n_{\rm e} \propto 1/M\dot{m}^{2}$, where $M$ is in physical units and $\dot{m}$ is the accretion rate in dimensionless units; \citealt{Svensson1994}). In addition to this, in recent years an interest has developed in understanding the potential systematic biases in relativistic X-ray reflection model parameters due to the assumption of suboptimal disk densities \citep{Garcia2016,Garcia2018b,Tomsick2018}. Indeed \cite{Tomsick2018} showed, via reflection modeling of Cyg~X-1, that low-density reflection models may lead to  spectral fits that are biased toward overabundances of iron in BHB accretion disks. 


More recently, attention has been called to the consistency of relativistic reflection modeling results that imply minimal disk truncation during the mid-to-bright hard X-ray spectral states ($L_{\rm X} \sim 0.01\mbox{--}0.1~L_{\rm Edd}$) of BHBs \citep{Zdziarski2020a,Zdziarski2021a}. Arguments were put forward, for example, relating to the physicality of the $R_{\rm in}$ constraint found by \citetalias{Garcia2018a}. \cite{Zdziarski2021a}---\citetalias{Zdziarski2021a} hereafter---principally argued that (i) the low-energy ($<1$~keV) coverage provided by simultaneous {\it Neil Gehrels Swift Observatory} (\swift) X-ray Telescope (XRT; \citealt{Burrows2005}) observations during the hard state require a disk blackbody flux consistent with a much more highly truncated disk ($R_{\rm in}>60~R_{\rm g}$); (ii) the disk heating, and thus associated effective temperature, caused by the irradiative flux impinging on the inner disk places a physical limit on $R_{\rm in}$ ($>100~R_{\rm g}$). Whilst these arguments were made specifically in the case of the hard state of \j1752, \cite{Zdziarski2020a} first put forward argument (ii) regarding the constraints from irradiation as a general comment on BHB hard states. We argue that this conundrum needs to be exhaustively tested by applying high-density reflection models to the high signal-to-noise observations of \j1752. 

In this paper we perform full, high-density reflection modeling of simultaneous \swift-XRT and \rxte\ observations of \j1752 during the hard state plateau that was the focus of both \citetalias{Garcia2018a} and \citetalias{Zdziarski2021a}; we note that the analysis of \citetalias{Garcia2018a} included only the \rxte\ data. We explore a range of potential geometrical setups for the disk-corona connection---including a rigorous test of the two-corona hypothesis proposed by \citetalias{Zdziarski2021a}---and discuss the pitfalls of the opposing model scenarios. {\it Our primary goal is to constrain the inner disk radius, $R_{\rm in}$, and explore its model-dependent systematics. } 

The structure of this paper is as follows. In Section~\ref{sec:data} we describe the \swift-XRT and \rxte\ data selection and reduction procedure. In Section~\ref{sec:swift_modeling} we discuss our spectral fits to the combined {\it Swift}-XRT spectrum and the resultant disk emission constraints. In Section~\ref{sec:modeling} we describe our full X-ray spectral modeling setup and present the results of our reflection modeling. In Section~\ref{sec:discussion} we discuss the implications of our results and give our conclusions. 

\section{Observations}\label{sec:data}

\begin{figure}
\includegraphics[width=\linewidth]{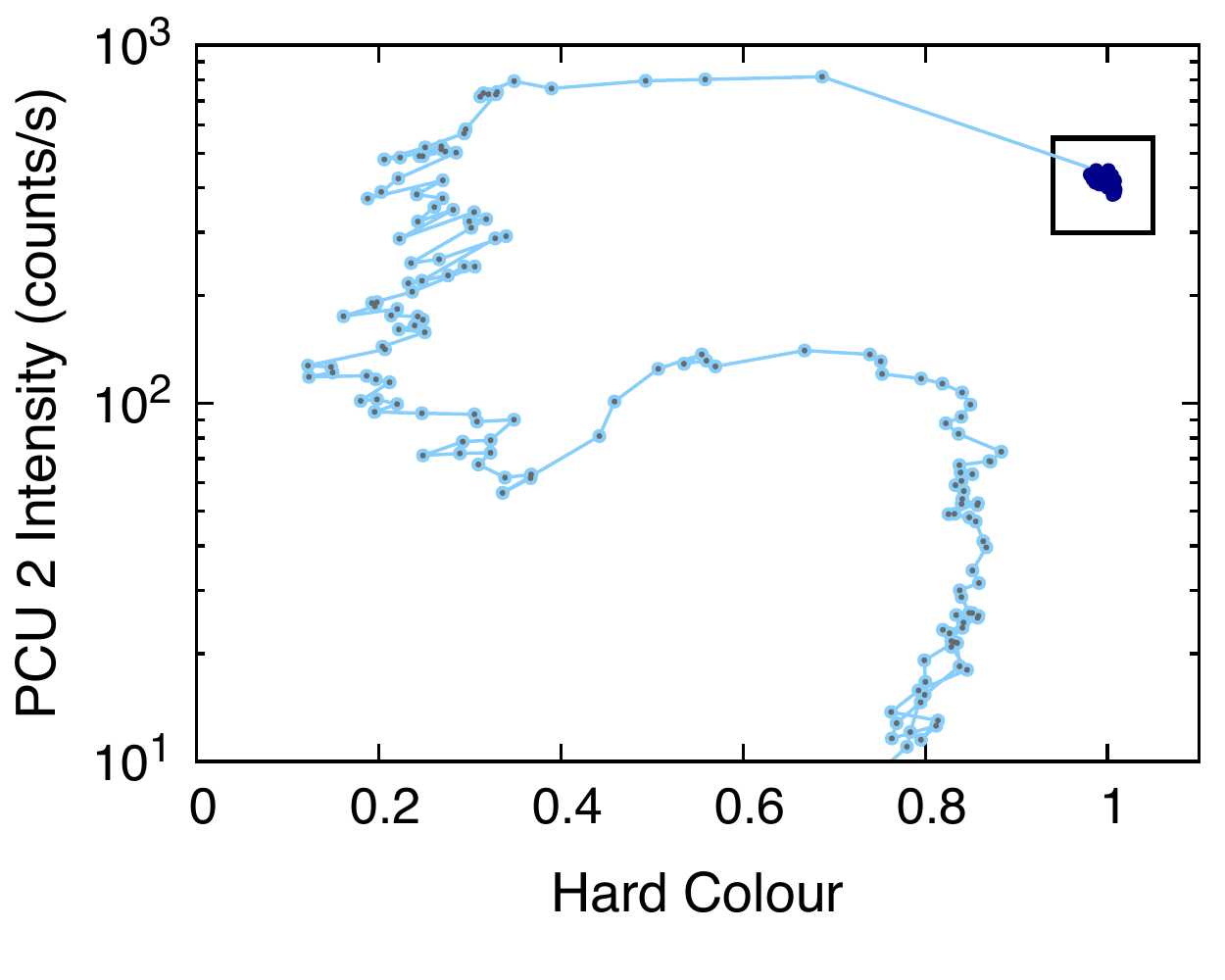}
\includegraphics[width=\linewidth]{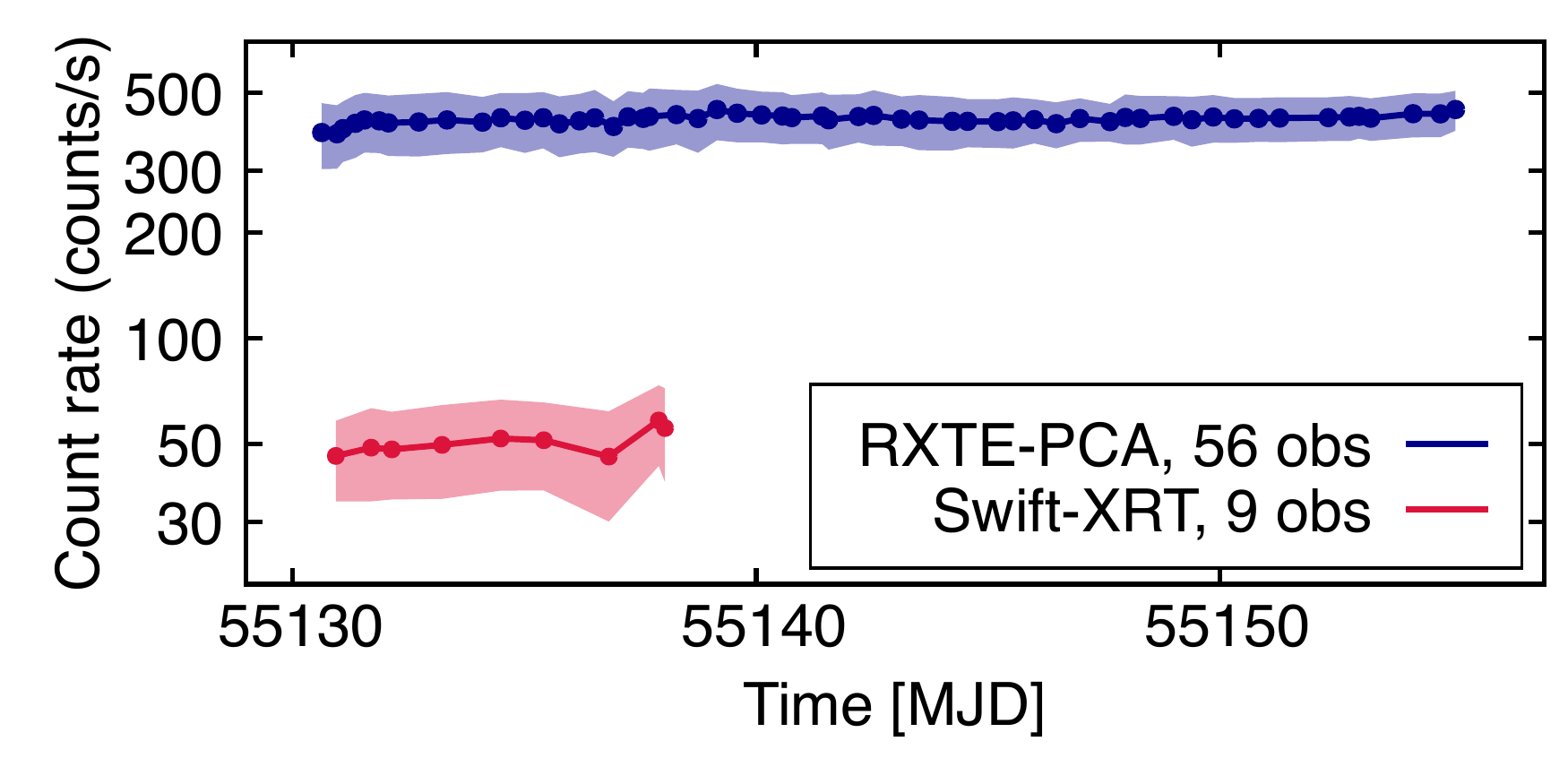}
\caption{Full hardness intensity diagram (top) and hard-state-only lightcurve (bottom) of the $2009\mbox{--}2010$ outburst of \j1752. The box (top panel) indicates the hard state plateau that is the focus of this analysis, the bottom light curves show data from that plateau region. }
\label{fig:data}
\end{figure}

All the \rxte\ data from observations of \j1752 during its 2010 outburst are publicly available on the \rxte\ archive via the HEASARC\footnote{\url{https://heasarc.gsfc.nasa.gov/docs/archive.html}} (High Energy Astrophysics Science Archive Research Center). The full outburst's hardness-intensity diagram (HID) and lightcurve are shown in Figure~\ref{fig:data}. The blue highlighted points display the \rxte\ data we select from the full outburst. We focused on {\it 56} observations taken as the source underwent a remarkably steady `plateau' in its hard spectral state. The light-curve in the lower panel shows that the Proportional Counter Array (PCA; \citealt{Jahoda_2006a}) count rate remains almost unchanged, staying within the level of the short term rms variability. We also show the simultaneous \swift-XRT observed count rate is consistent with being constant within the short term variability. 

\subsection{\rxte: PCA and HEXTE}
\label{subsec:rxte}
We extract the PCA spectra after removing data lying within $10$~min of the South Atlantic Anomaly (SAA). We use data from proportional counter unit (PCU) 2, due to its superior calibration and extensive coverage (all PCA exposures include PCU~2 data). We correct the PCU~2 data using the publicly available tool {\tt pcacorr} \citep{Garcia2014b}, and introduce~$0.1$\% systematic errors to all channels. The level of systematic errors to impose on the data is based upon the direction of \cite{Garcia2014b}, showing that the reduction in residuals achieved by the {\tt pcacorr} tool allows one to lower the assumed systematics to $\sim0.1$\%. We ignore PCU~2 data in channels $1\mbox{--}4$, and beyond $45$~keV. We extract the High Energy X-ray Timing Experiment (HEXTE) data from cluster B only because cluster A failed prior to the outburst of \j1752. This is all consistent with the reduction/analysis of \citetalias{Garcia2018a}. 

We then combine all the \rxte\ observations in identical fashion to \citetalias{Garcia2018a}. These data consist of all the spectra taken during the stable hard state plateau (purple data in Figure~\ref{fig:data}), combined to form one spectrum with $\sim100$~million and $\sim5$~million PCA and HEXTE~B counts respectively. The {\tt pcacorr} \citep{Garcia2014b} and {\tt hexbcorr} \citep{Garcia2016} tools were then applied to each respectively to reduce the systematics. We treat the data identically to \citetalias{Garcia2018a}: we add $0.1$~\% systematics to the PCA, and bin the HEXTE B spectrum by factors of 2, 3, and 4 in the $20\mbox{--}30$~keV, $30\mbox{--}40$~keV, and $40\mbox{--}250$~keV bands respectively. We then exclude the first 4 channels of the PCA, and include data up to $45$~keV, and include HEXTE data in the $20\mbox{--}140$~keV range.

\subsection{\swift-XRT}
\label{subsec:swift}
We also include the simultaneous \swift-XRT data in our analysis. There were a total of 9 \swift-XRT snapshots during the hard state plateau shown in Figure~\ref{fig:data}, taken in Window Timing (WT; \citealt{Hill2004}) mode. We extract all 9 observations identically, adopting a circular region of radius 47" centered on the source, and a background annulus region between 212'' and 259'', and selecting only grade 0 events. We generate the ancillary response files using the {\tt xrtmkarf} task, and used response file {\tt swxwt0s6\_20090101v015.rmf}. We set the background file scaling factor to 2 in the source spectrum file, and set the background scaling factor to 1 in the background file. A complete analysis of the individual PCA observations that were simultaneous with \swift-XRT pointings is shown in Section~\ref{subsec:individual}. 

We also combine all 9 \swift-XRT observations into a single spectrum using the FTOOL {\tt addspec}, which itself generates an ARF for the combined spectrum. The resultant combined \swift-XRT spectrum matches closely with that modeled by \citetalias{Zdziarski2021a}. We group the final combined data with a signal-to-noise ratio of 50. All analysis of the co-added \rxte\ and \swift-XRT spectra is shown in Section~\ref{subsec:combined}. 

\section{{\it Swift}-XRT: Modeling the Disk Spectrum}
\label{sec:swift_modeling}

\citetalias{Zdziarski2021a}, extending the work of \citetalias{Garcia2018a}, derived inner disk radius ($R_{\rm in}$) constraints during the hard state of \j1752 in part by modeling the combined \swift-XRT spectrum in the $0.55\mbox{--}6$~keV band, after applying $1\%$ systematic errors to all channels. Adopting a Comptonized disk spectral model given by {\tt crabcorr*TBabs*thcomp(diskbb)}, where {\tt thcomp} is a novel thermal Comptonization routine extended to mildly relativistic electron temperatures \citep{Zdziarski2020c}, they found a disk normalization of $N_{\rm disk}=0.9^{+1.6}_{-0.5}\times10^{6}$. Here the {\tt diskbb} component represents the multitemperature blackbody disk, assuming a non-zero torque boundary condition \citep{Mitsuda1984}, and the {\tt crabcorr} component is an instrumental cross-calibration model benchmarked by Crab observations, with a slope offset ($\Delta\Gamma$) and normalization ($N$) parameter \citep{Steiner2010}. \citetalias{Zdziarski2021a} noted that the choice of interstellar atomic abundances assumed in the {\tt TBabs} absorption calculation affects the disk normalization, with the abundances of \cite{Wilms2000} yielding slightly higher values than those of \cite{Anders1989}. We note, however, that the differences are minor when fitting to the \swift-XRT spectrum (the same cannot be said for the high-sensitivity PCA data). Furthermore, when fitting to such a narrow band of \swift-XRT data, the choice of disk model is a key driver of constraints on $N_{\rm disk}$ and thus $R_{\rm in}$. Throughout our analysis all spectral fits were performed using Xspec v12.10.1s. 

\begin{figure}
\centering
\includegraphics[width=\linewidth]{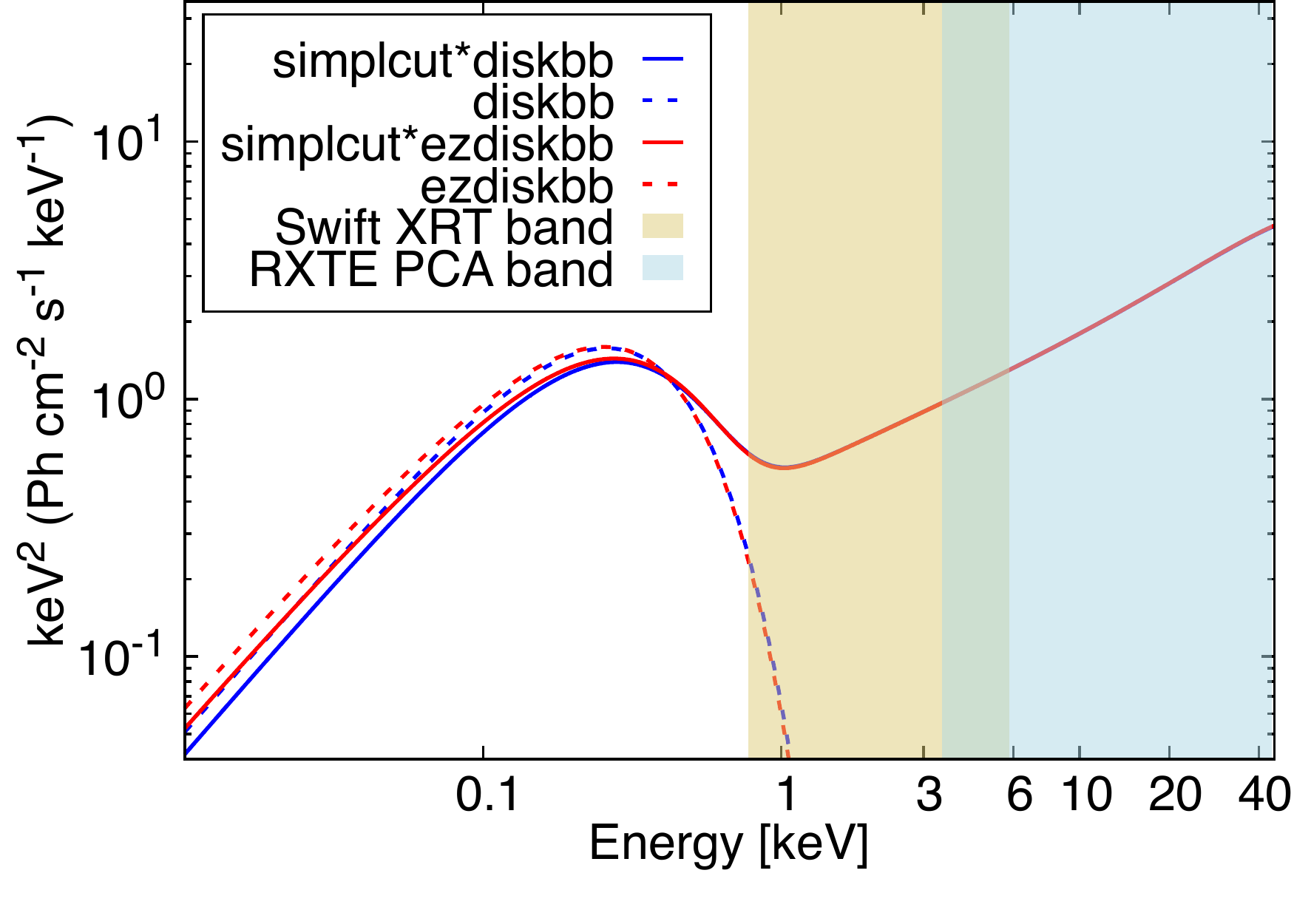}
\includegraphics[width=\linewidth]{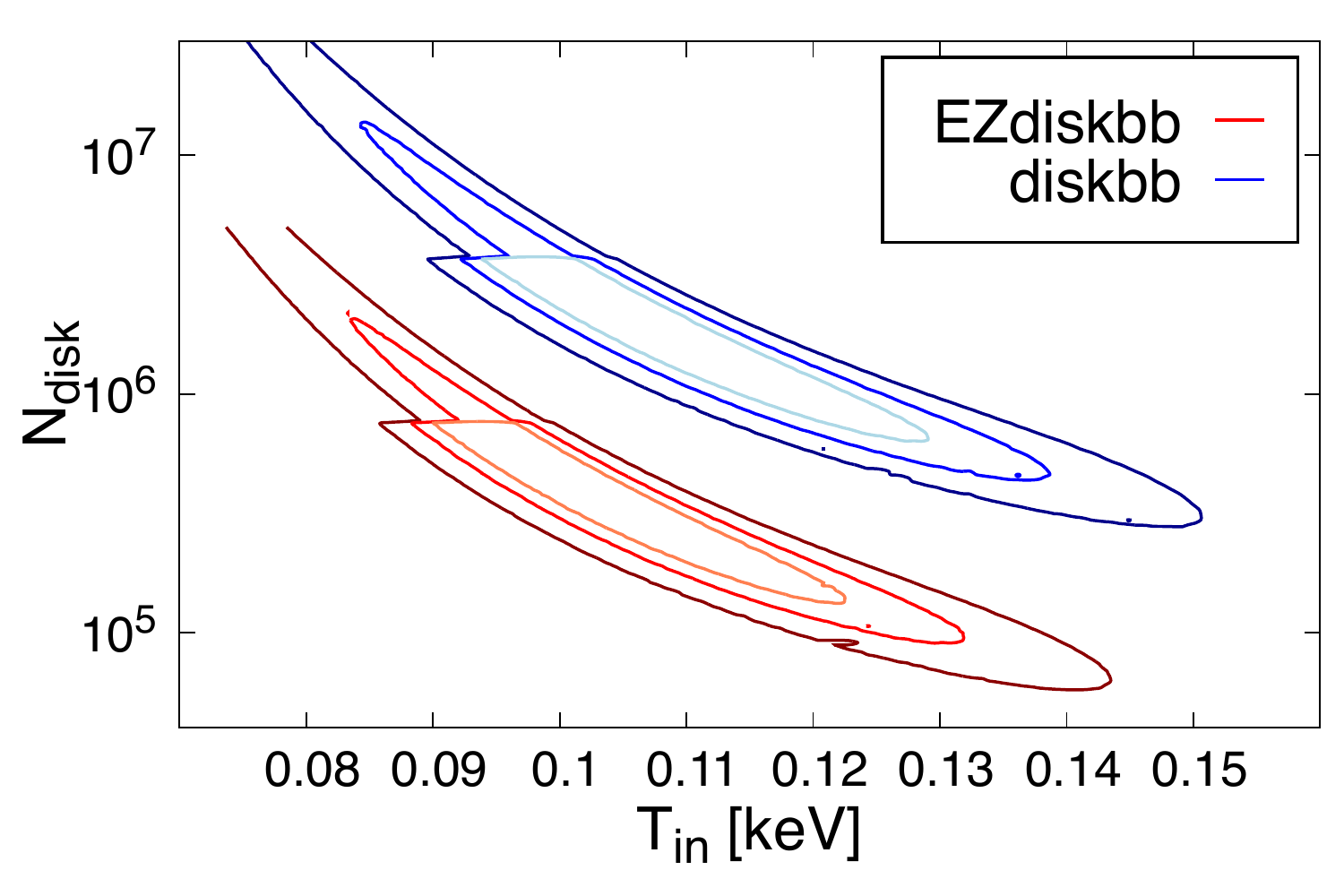}
\caption{A comparison of Comptonized disk continuum fits to the combined \swift-XRT spectrum of \j1752. {\bf Top}: the {\tt diskbb} (blue) and {\tt ezdiskbb} (red) unabsorbed models, comptonized by the {\tt simplcut} convolution routine, showing the similarity in their spectral shapes. The gold/blue shaded regions indicate the observable bands of \swift-XRT/\rxte-PCA respectively. The intrinsic disk emission only begins to dominate below $1$~keV. {\bf Bottom}: Contours of disk normalization, $N_{\rm disk}$, against inner disk temperature, $T_{\rm in}$, showing the comparison of constraints between {\tt diskbb} and {\tt ezdiskbb}. }
\label{fig:xrt_fits}
\end{figure}

We fit the same combined \swift-XRT spectrum---including the 1\% systematic errors---comparing the models {\tt diskbb} \citep{Mitsuda1984} and {\tt ezdiskbb} \citep{Zimmerman2005} as the disk continuum inputs for Comptonization. The difference between these two disk models lies in the radial temperature profile: {\tt diskbb} assumes a non-zero torque boundary condition at the inner edge of the disk, whereas {\tt ezdiskbb} adopts a zero-torque boundary condition \citep{SS1973,NT1973,Zimmerman2005}. \cite{Zimmerman2005} show that assuming a zero torque boundary condition results in systematically lower constraints on $T_{\rm in}$ and $R_{\rm in}$ by factors of roughly $5\%$ and $50\%$ respectively.  

We also include two additional Gaussian components in our spectral model to fit remaining systematic residuals around the $\sim2$~keV region (see, e.g., \citealt{MunozDarias2010}). Following the procedure of \citetalias{Zdziarski2021a}, we fix the Gaussian energies at $1.7$~keV and $2.3$~keV, and allow their widths ($\sigma$) and normalizations to vary freely.

Our total model is {\tt crabcorr*Tbabs(simplcut$\otimes$DISK + gau + gau)}, where {\tt DISK} is substituted for by each disk component in kind. The {\tt simplcut} component is a model for self-consistent Comptonization of the disk component, and acts as a convolution model \citep{Steiner2017}, with the often adopted {\tt nthComp} Comptonization spectrum \citep{Zdziarski1996,Zycki1999} used to produce the powerlaw component. Since {\tt simplcut} adopts {\tt nthComp} as its Comptonization continuum, it differs from {\tt thcomp}, which serves as an update to {\tt nthComp} in order to handle electron temperatures in the transrelativistic regime $\ge100$~keV \citep{Zdziarski2020c,Niedzwiecki2019}. As such, we expect general agreement between our approach and that of \citetalias{Zdziarski2021a} for lower electron temperatures. We therefore fixed $kT_{\rm e}=30$~keV since we are only interested in fitting the \swift-XRT spectrum up to $10$~keV. The best fit model in each case is shown in Figure~\ref{fig:xrt_fits}a. The gold shaded region shows the $0.55\mbox{--}6$~keV XRT band. One notices first of all that the disk component is only barely captured just below $1$~keV. There are minor differences between the two disk components, but significant differences in the disk normalization constraints, as shown in the $N_{\rm disk}$ vs $T_{\rm in}$ contour plot in Figure~\ref{fig:xrt_fits}b.

Using the standard formula of \cite{Kubota1998}, one can derive a color-corrected inner disk radius value from the disk normalization,

\begin{equation}
 R_{\rm in} = (D_{10}^2 N_{\rm disk} / \cos{i})^{1/2} \kappa^2~{\rm km}\,\,\, , 
 \label{eq:diskbb}
 \end{equation}
 
 where $D_{10}$ is the source distance in units of 10~kpc, $i$ is the disk inclination, and $\kappa$ is the total color correction factor, with a predicted range of $\kappa=1\mbox{--}2$ in broad terms. Each of these variables contain significan0t uncertainties, including the black hole mass, $M_{\rm BH}$, which allows us to translate $R_{\rm in}$ into gravitational units. Assuming ranges in distance of $D\sim3\mbox{--}6$~kpc, black hole mass $M_{\rm BH}=8\mbox{--}11~M_{\odot}$, $i=10^{\circ}\mbox{--}49^{\circ}$, and $\kappa=1.2\mbox{--}1.7$, with maximal spin, and the $90\%$ range of $N_{\rm disk} = 6\mbox{--}440\times10^5$ and $N_{\rm disk} = 2\mbox{--}17\times10^5$ for {\tt diskbb} and {\tt ezdiskbb} fits respectively, we find $R_{\rm in} \sim 20\mbox{--}970~R_{\rm ISCO}$ and $R_{\rm in} \sim 10\mbox{--}190~R_{\rm ISCO}$ in each case. This gives an aggregate allowed range, taking into account the different model assumptions, of $R_{\rm in} \sim 10\mbox{--}970~R_{\rm ISCO}$. 
 
 {\it So, based on modeling of the \swift-XRT spectrum alone, the $90\%$ limits on $R_{\rm in}$ range over almost two orders of magnitude. In other words, the direct disk emission is poorly constrained, with more room for discussion on whether the disk can be close to the ISCO in this stable hard state plateau, or is more likely to be truncated. }

\section{Simultaneous Modeling} \label{sec:modeling}

We perform a more in depth analysis of the combined \swift-XRT and \rxte\ data, covering the roughly month-long stable hard state (plateau; Figure~\ref{fig:data}). Here we sought to investigate the geometry of \j1752 in this stable hard state configuration, in lieu of the conflicting models presented by \citetalias{Garcia2018a} and \citetalias{Zdziarski2021a}. 

In contrast to both \citetalias{Garcia2018a} and \citetalias{Zdziarski2021a}, we explore reflection modeling of the high signal-to-noise combined hard state \swift-XRT/\rxte\ spectrum with high disk density reflection models. In the Subsections that follow (\ref{subsec:individual} and \ref{subsec:combined}) we show results of modeling analysis of both the individual strictly simultaneous \swift-XRT and \rxte\ observations in the hard state, as well as the total combined \swift-XRT and \rxte\ spectrum (similarly to the analysis presented in \citetalias{Zdziarski2021a}).


\subsection{Individual observations}
\label{subsec:individual}
The analyses of \citetalias{Garcia2018a} and \citetalias{Zdziarski2021a} were performed using combined PCA-HEXTE, and in the latter paper, \swift-XRT spectra, such that model constraints are maximized by high signal-to-noise in tandem with broadband energy coverage. It is still worth, however, exploring any potential modeling biases that could have been caused by combining data across multiple observations of the source. We therefore begin by analyzing all the 9 individual, strictly simultaneous \swift-XRT and PCA observations. We group the individual PCU~2 spectra at a signal-to-noise ratio of 4, which achieves sufficient oversampling of the source counts to outweigh the background at high energies. We group the individual \swift-XRT spectra at a signal-to-noise ratio of 10, this time applying no systematic errors, and include the data between $0.5\mbox{--}10$~keV because the lower number of counts in individual spectra (as opposed to combined spectra) means that statistical (as opposed to systematic) errors dominate.


We test the capability of these individual spectra to provide constraints on $R_{\rm in}$, which is primarily driven by the broadness of the Fe K line as measured by the PCA. We adopt the model {\tt crabcorr*Tbabs(simplcut$\otimes$ezdiskbb+relxillCp)}. We choose a low-density reflection model ($n_{\rm e}=10^{15}~{\rm cm^{-3}}$) in order to make a direct comparison with the constraints found by both \citetalias{Garcia2018a} and \citetalias{Zdziarski2021a}. We treat the {\tt crabcorr} correction model restrictively: the slope offsets are all set to values given by the instrument calibration as tabulated by \cite{Steiner2010}, such that $\Delta\Gamma=0.01$ and $N=1.097$ in the PCA, $\Delta\Gamma=0.01$ and $N={\rm free}$ in the HEXTE B spectrum, and $\Delta\Gamma=-0.04$ and $N={\rm free}$ in the XRT spectrum. This approach to handling the cross-calibration of the respective instruments avoids model biases that could arise from over-corrections, given the complexity of the models being applied. We adopt this setup throughout our analysis. The final component, {\tt relxillCp}, is the relativistic reflection model, a flavor of the {\tt relxill} suite of models \citep{Garcia2014,Dauser2014} that assumes the irradiating continuum is a Comptonization spectrum, treated identically to {\tt nthComp}. 

\begin{figure}
\centering
\includegraphics[width=0.98\linewidth]{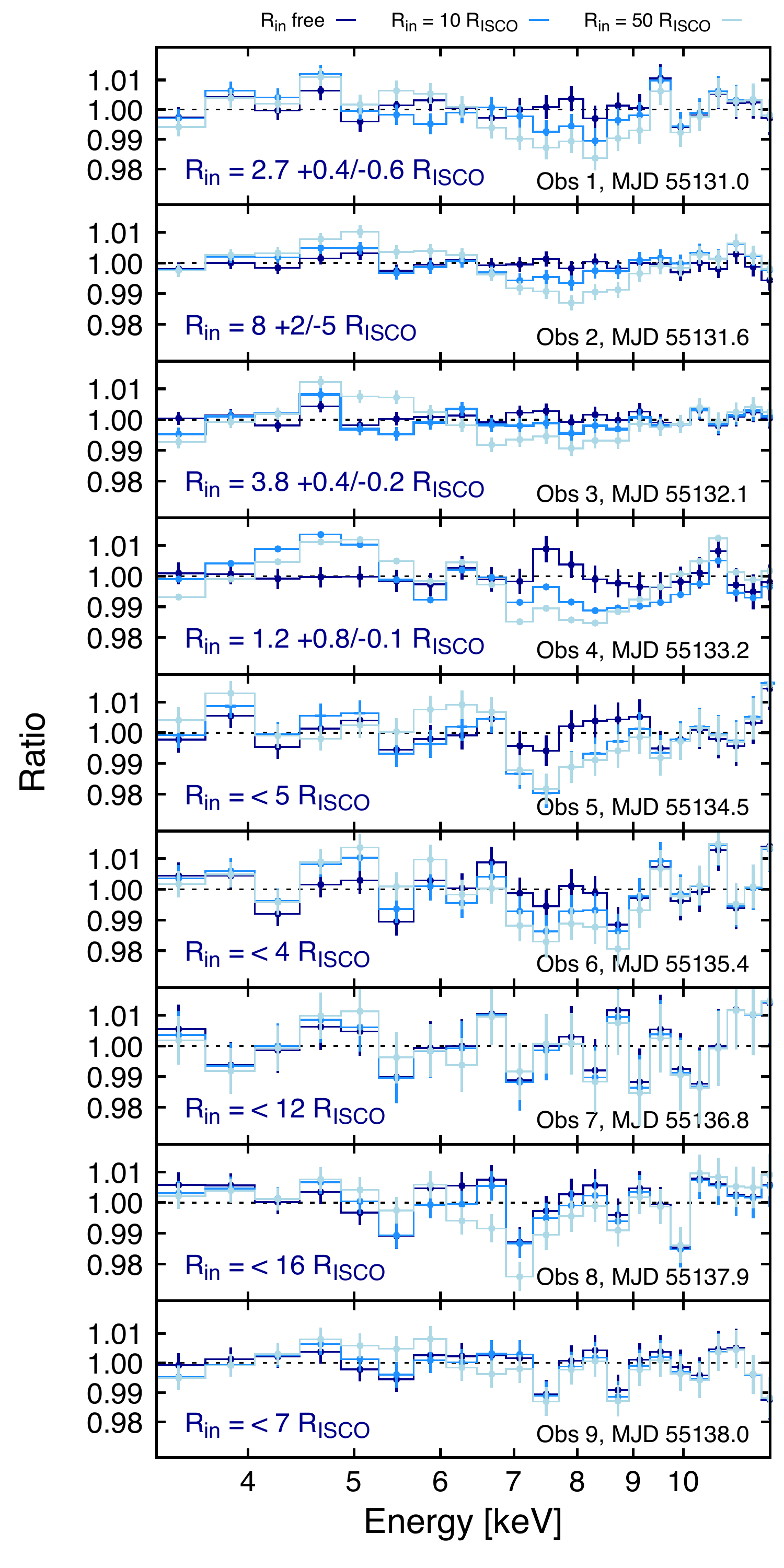}
\caption{Progression of ratio (data/model) residuals in fits to the hard state plateau simultaneous individual \rxte\ and \swift-XRT observations. We show only the PCA data in the $3\mbox{--}12$~keV band to emphasize the residuals around the Fe K line region. Fits were performed with $R_{\rm in}$ free (dark blue), $R_{\rm in}=10~R_{\rm ISCO}$ (blue), and $R_{\rm in}=50~R_{\rm ISCO}$ (light blue). The $R_{\rm in}$ constraints from fitting the parameter freely are shown in each case in dark blue}.
\label{fig:chi_rin}
\end{figure}

We fit each simultaneous $0.5\mbox{--}45$~keV XRT/PCA spectrum adopting three different treatments of the inner disk radius parameter, $R_{\rm in}$ free, $R_{\rm in}=10~R_{\rm ISCO}$, and $R_{\rm in}=50~R_{\rm ISCO}$---note that here we refer only to the $R_{\rm in}$ parameter of the {\tt relxillCp} component, and thus the constraints that follow relate to the reflection spectrum. Figure~\ref{fig:chi_rin} shows the ratio (data/model) residuals resulting from each of these fits, where we limit the energy range to $3\mbox{--}12$~keV in the plots for the sake of clarity. It is clear that consistently across all 9 observations, to varying degrees dependent upon signal-to-noise, truncating the disk results in poorer fits to the line region---in all cases in which $R_{\rm in}$ is variable, it settles close to the ISCO, with varying degrees of certainty, again dependent upon signal-to-noise. It is important to note that this applies to both the redward and blueward sides of the line emission; models assuming truncation fail to capture both the broadened red wing and the smeared blue edge correctly. It is therefore beyond any doubt that if we assume the coronal continuum emission can be well represented by a single-temperature thermal Comtponization spectrum ({\tt simplcut$\otimes$ezdiskbb}), reflection is occurring in the inner $10~R_{\rm g}$ of the accretion flow. 

\citetalias{Zdziarski2021a} have suggested, however, that the coronal continuum is likely more complex in shape than captured by thermal Comptonization model. They argue that the fast variability of the coronal emission, as well as additional coronal regions at variable temperatures and optical depths, would lead to a more curved irradiating continuum. If this is the case, the observed reflection features could be somewhat subsumed by the direct continuum emission, leaving residuals that can be explained by a less smeared reflection spectrum from a truncated disk. We therefore test this exhaustively in the Section~\ref{subsubsec:hd-hard}.

\subsection{Combined observations}
\label{subsec:combined}

\citetalias{Garcia2018a} performed an exhaustive reflection modeling analysis of the combined \rxte\ (PCA \& HEXTE) observations of \j1752. They did not, however, include the contemporaneous \swift-XRT data in their analysis. As such, \citetalias{Garcia2018a} only explored the hard powerlaw continuum in the hard state, along with the reflection spectrum. \citetalias{Zdziarski2021a} expanded upon this analysis by including a summed \swift-XRT spectrum (all 9 snapshots summed using the {\tt addspec} FTOOL), allowing constraints on the disk spectrum, visible as a soft excess in the \swift-XRT spectrum below $\sim1$~keV. Here we begin from the analysis of \citetalias{Garcia2018a} and show a step-by-step analysis of the comparative results when including the \swift-XRT spectrum in our full dataset, as described in full in Section~\ref{sec:data}, before moving onto more novel high-density reflection modeling of the data.

\subsubsection{Low-density reflection modeling}
\label{subsubsec:ld-hard}

First we adopted Model~$1$ implemented by \citetalias{Garcia2018a}: {\tt crabcorr*TBabs(relxillCp + xillverCp + gau + gau)}, in which the assumed black hole spin is maximal $a_{\star}=0.998$, the inner disk radius $R_{\rm in}$ varies freely, and the disk density is low ($n_{\rm e}=10^{15}~{\rm cm^{-3}}$). The two additional Gaussians are included due to the appearance of residuals at $29.8$~keV (the energy of a $^{241}{\rm Am}$ radioactive emission line; \citealt{Jahoda_2006a}) and $40\mbox{--}45$~keV (which is of unknown origin). \citetalias{Garcia2018a} include these two Gaussian lines in their fits, the latter of which settles on $\sim43.2$~keV. We therefore fix the energies and widths of each line at $28.9$~keV/$0.1$~keV and $43.2$~keV/$0.1$~keV respectively. We adopt the solar abundances of \cite{Wilms2000} and atomic cross-sections of \cite{Verner1996}. We achieve an identical fit to the combined PCA and HEXTE~B spectrum to that of \citetalias{Garcia2018a}.

\begin{figure}
\centering
\includegraphics[width=0.98\linewidth,trim={0 1.2cm 0 0}]{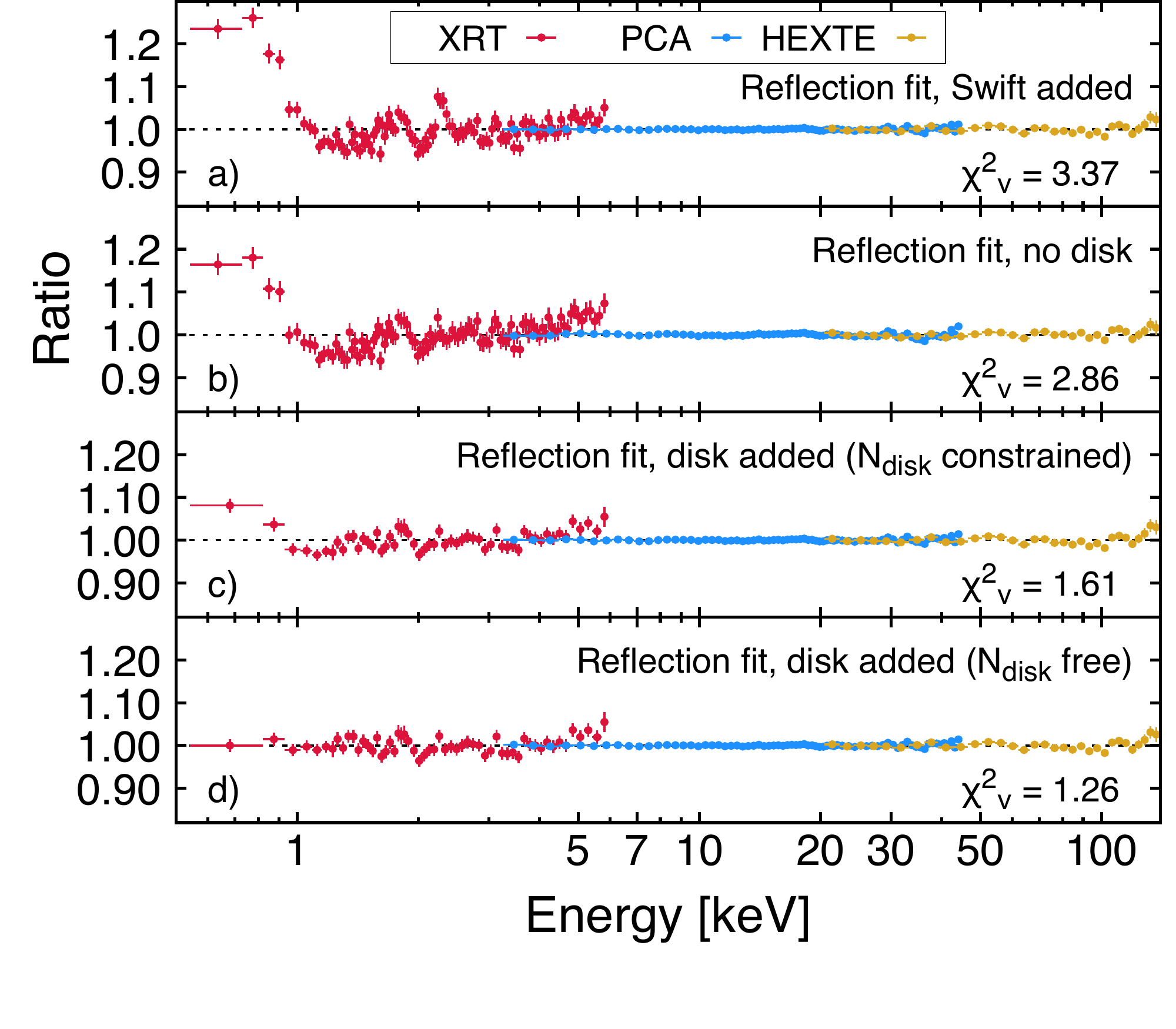}
\caption{Progression of ratios in fits to the hard state plateau combined \rxte\ and \swift-XRT data, based upon the results of \citetalias{Garcia2018a}. Panels in order are as follows: a) Model 1 of \citetalias{Garcia2018a} with the \swift-XRT spectrum added; b) the same model re-fit to the total broadband spectrum; c) disk added to the model, with the normalization constrained by the $R_{\rm in}$ value of the {\tt relxillCp} component; d) same as c), except the disk is free to vary in normalization.}
\label{fig:hard-progression}
\end{figure}

 We then include the combined \swift-XRT spectrum, as described in Section~\ref{sec:data}, and inspect the residuals. This is shown in panel a of Figure~\ref{fig:hard-progression}. One can see a soft excess below $1$~keV, a likely sign of the disk blackbody spectrum, regardless of the heating mechanism (irradiative or viscous/magnetic). We then attempt a fit to the full dataset, adopting only the {\tt crabcorr} parameters corresponding to the \swift-XRT data as additional free parameters, and allowing the same model parameters to vary as shown in the Model~$1$ results of \citetalias{Garcia2018a} (thus, no disk is included in the model; panel b of Figure~\ref{fig:hard-progression}). Panel b clearly shows that we cannot fit the full broadband spectrum with Model~$1$ of \citetalias{Garcia2018a}, i.e., a reflected power law model cannot explain the soft energy spectrum, assuming low disk density ($n_{\rm e}=10^{15}~cm^{-3}$). 
 
 Finally, we add a disk blackbody component to the model, in the form of {\tt ezdiskbb}, adopting two approaches. First we constrain the disk normalization in accordance with the $R_{\rm in}$ parameter of the {\tt relxillCp} component, by simply rearranging Equation~\ref{eq:diskbb}. We fold in the uncertainties in black hole mass, $M_{\rm BH}$, disk inclination, $i$, source distance, $D$, and color correction factor, $\kappa$, by introducing a variable that is dependent on all four quantities, $\Sigma \propto (M_{\rm BH}^2\cos i/D^2 \kappa^4)$. Our baseline is such that $\Sigma=1$ when $M_{\rm BH}=10~M_{\odot}$, $i=30^{\circ}$, $D=3.5~{\rm kpc}$, and $\kappa=1.7$, therefore $N_{\rm disk} = \Sigma M_{10}^2 \cos i_{30} / D_{3.5}^2 \kappa_{1.7}^4$. The introduction of the $M_{\rm BH}$ term follows simply from the size scaling, i.e., $R_{\rm in} \propto M_{\rm BH}$, and since $R_{\rm in}$ in {\tt relxillCp} is in units of $R_{\rm ISCO}$, there is a mass-dependence introduced to the absolute disk flux normalization. Adopting the ranges $M_{\rm BH}=8\mbox{--}11~M_{\odot}$, $i=10^{\circ}\mbox{--}49^{\circ}$, $D=3\mbox{--}6~{\rm kpc}$, and $\kappa=1.2\mbox{--}1.7$, leads to the limits $0.16\le\Sigma\le7.54$.
 
  We find a moderate improvement to the fit (panel c of Figure~\ref{fig:hard-progression}), and as one would expect, $\Sigma$ increases to its maximal value, corresponding to the lower limits on $i$ ($30^{\circ}$), $D$ ($3$~kpc) and $\kappa$ ($1.2$), and upper limit on $M_{\rm BH}$ ($11~M_{\odot}$). Residuals remain in the soft band below $1$~keV, however, on the order of $\sim10\%$. Second, we allow the disk normalization to vary independently. We find that including a disk component in this way greatly improved the fit to the data, showing a clear reduction in the residuals below $1$~keV, as shown in panel d of Figure~\ref{fig:hard-progression}. In all fits we let the hydrogen column density, $N_{\rm H}$, vary freely, and recover values close to $10^{22}~{\rm cm^{-2}}$. 

\begin{figure}
\centering
\vspace{-0.3cm}
\includegraphics[width=0.98\linewidth]{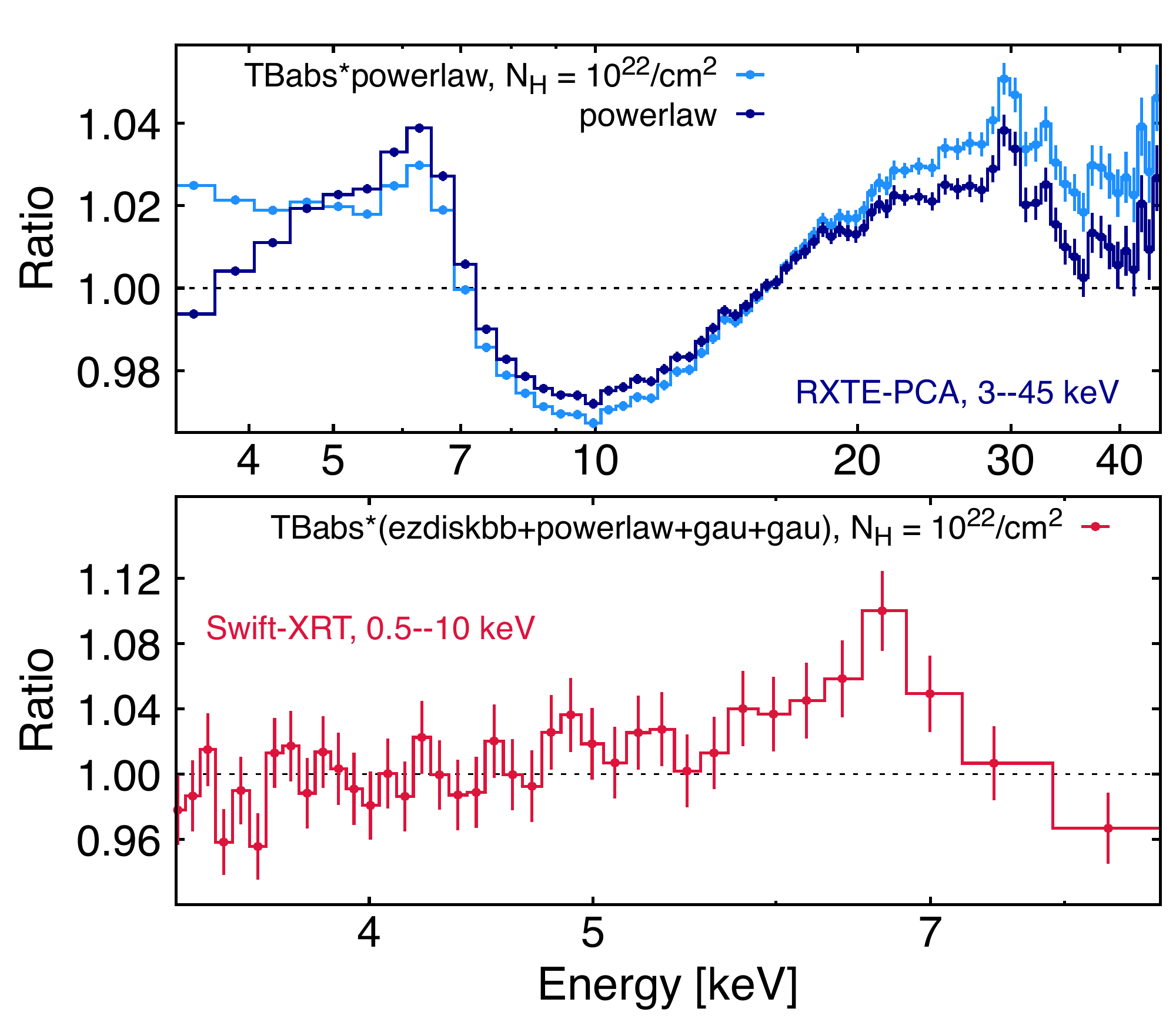}
\vspace{-5pt}
\caption{{\bf Top}: Ratios of the total combined PCA spectrum of the hard state observations of \j1752 fit with a {\tt powerlaw} model and absorbed model {\tt TBabs*powerlaw}. Despite the relatively coarse energy resolution of the PCA detectors ($\sim1~$keV at $6$~keV), the increased sensitivity of combining all 56 observations means the iron emission line displays enough complexity to distinguish the broadened component from an additional narrow line feature at $\sim6.4$~keV. {\bf Bottom}: Ratios of an absorbed disk$+$powerlaw fit to the \swift-XRT spectrum, showing only $3.3\mbox{--}9$~keV to emphasize the broad iron line region. Two Gaussian components are added to account for systematics at $1.7$~keV and $2.3$~keV.}
\label{fig:hard_state_powerlaw_ratios}
\end{figure}

Inclusion of a disk blackbody component instantly reveals the tension between the interpretations of \citetalias{Garcia2018a} and \citetalias{Zdziarski2021a} regarding the truncation of the accretion disk. Our inner disk radius constraint, as given by the relativistic reflection component, agrees well with that of \citetalias{Garcia2018a}, with $R_{\rm in} = 1.1 \pm 0.1~R_{\rm ISCO}$ (confidence intervals derived only from the covariance matrix, as is standard when performing the {\tt fit} command with Xspec). In other words, including a disk blackbody component to model out the soft-band residuals has no impact on the broad iron line constraints on $R_{\rm in}$. However, the preliminary (again taking just the analytically derived covariance matrix) disk parameter constraints give $T_{\rm in}=0.113\pm0.005$ and $N_{\rm disk} = 4 \pm 1\times10^5$. Using Equation~\ref{eq:diskbb}, and assuming a distance of $3\mbox{--}6$~kpc, black hole mass $M_{\rm BH}=8\mbox{--}11~M_{\odot}$, $i=10^{\circ}\mbox{--}49^{\circ}$, and $k=1.2\mbox{--}1.7$, with maximal spin, $N_{\rm disk} = 4\pm1\times10^5$ gives $R_{\rm in} \sim 10\mbox{--}100~R_{\rm ISCO}$. This again shows us that by invoking just the uncertainty on our estimate of $N_{\rm disk}$, and our ignorance regarding the system characteristics, the constraint on $R_{\rm in}$ from full broadband spectral modeling with the inclusion of the disk component is weak.

\begin{figure*}
\centering
\includegraphics[width=0.98\linewidth,trim={0 2cm 0 0}]{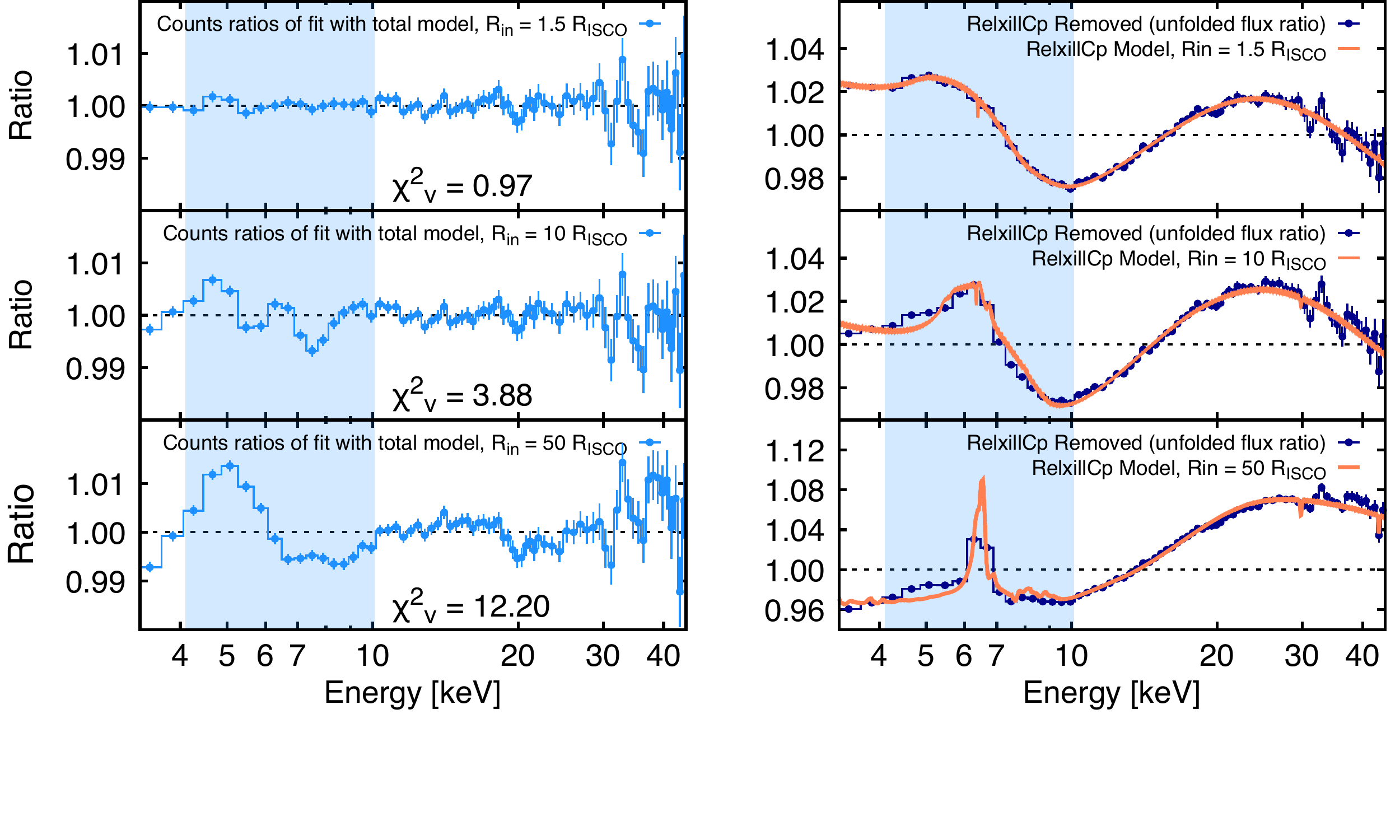}
\caption{Ratios of fits to the PCA data with the model {\tt TBabs(simplcut$\otimes$ezdiskbb+relxillCp+gau+gau+gau)}, where the first Gaussian component represents the narrow line component at $\sim6.4$~keV, the second two Gaussian components represent the systematic features at $29.8$~keV and $43.2$~keV. The left panels shows the residuals of fits to the spectrum with $R_{\rm in}$ free (top), $R_{\rm in}=10~R_{\rm ISCO}$ (middle) and $R_{\rm in}=50~R_{\rm ISCO}$ (bottom). The right hand panels show the ratio of the unfolded data to the model after removing the reflection component, with the model overlaid in coral. In all fits the hydrogen column density ($N_{\rm H}$) is allowed to vary. The shaded region indicates the $\sim4.1\mbox{--}10.1$~keV band (15 bins) modeled by \citetalias{Zdziarski2021a}. Some minor residuals can be seen in the right hand panels at the energies of the two model Gaussians ($29.8$~keV and $43.2$~keV) due to small normalization offsets when computing the unfolded data/model ratios.} 
\label{fig:hard_state_refl_ratios}
\end{figure*}

 We can get a clear understanding of the nature of the reflection features of \j1752 in its hard state by inspecting the PCA data, since the signal-to-noise is remarkably high in the Fe K region ($S/N\sim6000$ in the $4\mbox{--}8$~keV band). The top panel of Figure~\ref{fig:hard_state_powerlaw_ratios} shows the ratio residuals of a simple powerlaw fit to the combined PCA spectrum, both with and without interstellar absorption. The residuals show a strong, broad, and likely multi-component Fe K line, a broad Fe K edge, and strong Compton hump. When interstellar absorption is taken into account, the line region resembles a two-component narrow$+$broad reflection complex. The bottom panel of Figure~\ref{fig:hard_state_powerlaw_ratios} shows the ratio residuals of a disk$+$powerlaw fit to the \swift-XRT spectrum, where we emphasize the iron line region. Again, despite the significantly lower signal-to-noise ($S/N\sim200$ in the $4\mbox{--}8$~keV band), the broad line feature is detected.However, in order to understand whether or not these features are truly present, we must take a closer look at what the reflection model parameters reveal upon fitting to the data.

Figure~\ref{fig:hard_state_refl_ratios} shows a deeper look at fits to the combined PCA data (excluding the \swift-XRT spectrum for now), which dominates constraints on the reflection model. The high signal-to-noise, particularly in the $3\mbox{--}10$~keV band where the broad iron line and smeared edge features reside, places a strong constraint on the shape of the reflection spectrum, as was reported by \citetalias{Garcia2018a}. One can see that the relativistically smeared reflection component, represented by {\tt relxillCp}, contributes to the redward and blueward sides of the line region, despite the need for a narrow line component at $6.4$~keV. Allowing the hydrogen column density to vary cannot account for this signal, so the reflection spectrum is not simply degenerate with interstellar absorption at soft energies. In addition, the smeared iron edge at $E>7$~keV is poorly fit with a reflection component with relatively mild truncation ($R_{\rm in}=10~R_{\rm ISCO}$), and this only worsens for higher truncation ($R_{\rm in}=50~R_{\rm ISCO}$). This illustrates conclusively that when we take into account the full observable band of just one instrument (thus with no biases injected due to cross-calibration issues), relativistic reflection from a single Comtponization component cannot fit the data with high disk truncation. We note that limiting the energy range of the instrument simply circumvents the need to treat both the red wing of the broad line component and the blueward smeared edge simultaneously. This is also easy to understand by inspecting Figure~\ref{fig:hard_state_refl_ratios}. 

In Section~\ref{subsubsec:hd-hard} we explore other potential scenarios, such as multiple coronal IC components, to investigate the possibility of modeling out the reflection features under the assumption of a truncated disk. Our recent global BHB reflection modeling campaign has highlighted the importance of utilizing high-density reflection models when fitting for the reflection properties of BHBs \citep{Garcia2018b,Tomsick2018,JJiang2019,JJiang2019b,Connors2020,Connors2021}. The heating effects induced by increasing the density of the accretion disk \citep{Garcia2016} lead to a prominent spectral component at soft energies where the intrinsic disk emission appears. Thus particularly in hard spectral states, one must include these effects in order to fully explore the degeneracy between the intrinsic and reprocessed disk components (e.g., see \citealt{JJiang2019}).

 In the following section we explore full high-density reflection modeling of \j1752, and investigate the interplay between the reflected IC spectra and the intrinsic disk component. 

\subsubsection{High-density reflection modeling}
\label{subsubsec:hd-hard}


\def\mANH{$1.070^{+0.006}_{-0.007}$}
\def\mAga{$1.4438^{+0.0005}_{-0.0006}$}
\def\mAfsc{$0.237^{+0.001}_{-0.007}$}
\def\mAkTe{$50.7^{+0.8}_{-1.4}$}
\def\mATin{$0.1231^{+0.0005}_{-0.0007}$}
\def\mAndisk{$97.7^{+0.2}_{-3.4}$} 
\def\mAq{$>9.7$}
\def\mARin{$1.286^{+0.07}_{-0.006}$} 
\def\mAincl{$58.1^{+0.5}_{-1.0}$}
\def\mAa{$0.998$}
\def\mAlogxi{$4.411^{+0.003}_{-0.007}$}
\def\mAAfe{$>9.7$}
\def\mADensityOne{$15.9^{+2.0}_{-0.4}$} 
\def\mADensityTwo{$8^{+84}_{-6}$}
\def\mAnrelOne{$3.23^{+0.06}_{-0.04}$} 
\def\mAnrelTwo{$1.98^{+0.07}_{-0.05}$} 
\def\mALineE{$6.97\pm0.02$}
\def\mASigma{$<110$}
\def\mAStrength{$0.034\pm0.003$}
\def\mANHxstar{$10.7^{+0.1}_{-3.8}$}
\def\mAlogxixstar{$4.17^{+0.03}_{-0.18}$}
\def\mAvxstar{$2100^{+600}_{-900}$}
\def\mANccFPMA{$1^a$}
\def\mANccHEXTE{$0.918\pm0.001$} 
\def\mANccSwift{$0.853\pm0.003$}
\def\mAGauPCAoneN{$0.08^{+0.04}_{-0.05}$} 
\def\mAGauPCAtwoN{$0.14^{+0.04}_{-0.05}$} 
\def\mAGauSwiftoneN{$2.5^{+0.6}_{-0.4}$} 
\def\mAGauSwifttwoN{$3.5^{+1.1}_{-0.8}$} 
\def\mAGauSwiftoneSigma{$<0.07$} 
\def\mAGauSwifttwoSigma{$0.11^{+0.04}_{-0.03}$} 
\def\mAchi{$243$}
\def\mAnu{$191$}
\def\mAchired{$1.27$}

\def\mATwoNH{$1.100^{+0.005}_{-0.007}$}
\def\mATwoga{$1.468^{+0.002}_{-0.004}$}
\def\mATwofsc{$0.27^{+0.01}_{-0.06}$}
\def\mATwokTe{$48.9^{+0.2}_{-0.4}$}
\def\mATwoTin{$0.124\pm0.001$}
\def\mATwondisk{$106^{+43}_{-3}$} 
\def\mATwoq{$>9.5$}
\def\mATwoRin{$1.33^{+0.17}_{-0.06}$} 
\def\mATwoincl{$55^{+1}_{-5}$}
\def\mATwoa{$0.998$}
\def\mATwologxi{$4.32^{+0.01}_{-0.02}$}
\def\mATwoAfe{$9.71^{+0.07}_{-1.57}$}
\def\mATwoDensityOne{$20^a$} 
\def\mATwoDensityTwo{$8^{+84}_{-6}$}
\def\mATwonrelOne{$2.26^{+0.02}_{-0.10}$} 
\def\mATwonrelTwo{$2.6\pm0.01$} 
\def\mATwoLineE{$6.97\pm0.02$}
\def\mATwoSigma{$<110$}
\def\mATwoStrength{$0.034\pm0.003$}
\def\mATwoNHxstar{$10.7^{+0.1}_{-3.8}$}
\def\mAlogxixstar{$4.17^{+0.03}_{-0.18}$}
\def\mAvxstar{$2100^{+600}_{-900}$}
\def\mANccFPMA{$1^a$}
\def\mATwoNccHEXTE{$0.918\pm0.001$} 
\def\mATwoNccSwift{$0.857\pm0.003$}
\def\mATwoGauPCAoneN{$0.15^{+0.05}_{-0.04}$} 
\def\mATwoGauPCAtwoN{$0.12^{+0.05}_{-0.04}$} 
\def\mATwoGauSwiftoneN{$2.5^{+0.5}_{-0.6}$} 
\def\mATwoGauSwifttwoN{$4.6\pm0.9$} 
\def\mATwoGauSwiftoneSigma{$<0.07$} 
\def\mATwoGauSwifttwoSigma{$0.13^{+0.04}_{-0.03}$} 
\def\mATwochi{$254$}
\def\mATwonu{$192$}
\def\mATwochired{$1.32$}

\def\mBNH{$0.823^{+0.004}_{-0.008}$}
\def\mBga{$1.324^{+0.002}_{-0.001}$}
\def\mBfsc{$<0.04$}
\def\mBkTe{$130.1^{+0.9}_{-1.1}$}
\def\mBTin{$0.167^{+0.003}_{-0.001}$}
\def\mBndisk{$8.8^{+1.5}_{-0.9}$} 
\def\mBSIGMA{$6.8\mbox{--}7.5$} 
\def\mBRin{$2.01^{+0.09}_{-0.11}$} 
\def\mBq{$3.74^{+0.50}_{-0.07}$} 
\def\mBincl{$<18$}
\def\mBa{$0.998$}
\def\mBlogxi{$4.251\pm0.003$}
\def\mBAfe{$5.00\pm0.04$}
\def\mBDensityOne{$17.52^{+0.14}_{-0.08}$}
\def\mBDensityTwo{$6.2^{+19.5}_{-0.3}$}
\def\mBnrelOne{$10.91^{+0.07}_{-0.57}$} 
\def\mBnrelTwo{$3.6^{+0.1}_{-0.2}$} 
\def\mBLineE{$6.97\pm0.02$}
\def\mBSigma{$130^{+30}_{-40}$}
\def\mBStrength{$0.046\pm0.005$}
\def\mBNHxstar{$10.8^{+0.2}_{-3.1}$}
\def\mBlogxixstar{$4.09^{+0.04}_{-0.04}$}
\def\mBvxstar{$2100^{+600}_{-600}$}
\def\mBNccFPMA{$1^a$}
\def\mBNccHEXTE{$0.917\pm0.001$}
\def\mBNccSwift{$0.859\pm0.004$}
\def\mBGauPCAoneN{$0.11\pm0.05$} 
\def\mBGauPCAtwoN{$0.14\pm0.05$} 
\def\mBGauSwiftoneN{$3.3^{+0.8}_{-0.7}$} 
\def\mBGauSwifttwoN{$6\pm1$} 
\def\mBGauSwiftoneSigma{$0.06\pm0.03$} 
\def\mBGauSwifttwoSigma{$0.13^{+0.04}_{-0.03}$} 
\def\mBchi{$318$}
\def\mBnu{$191$}
\def\mBchired{$1.66$}

\def\mCNH{$1.14^{+0.05}_{-0.04}$}
\def\mCgaOne{$1.36\pm0.02$}
\def\mCgaTwo{$2.0\pm0.03$}
\def\mCfsc{$0.1268^{+0.0009}_{-0.0048}$}
\def\mCkTeOne{$97.7^{+5.4}_{-0.1}$}
\def\mCkTeTwo{$>100$}
\def\mCTin{$0.12\pm0.01$}
\def\mCndisk{$140^{+150}_{-70}$} 
\def\mCq{$3.69^{+0.02}_{-0.06}$} 
\def\mCRinOne{$1.47^{+0.22}_{-0.06}$} 
\def\mCRinTwo{$>60$} 
\def\mCincl{unconstrained}
\def\mCa{$0.998$}
\def\mClogxiOne{$4.47\pm0.05$}
\def\mClogxiTwo{$0^a$}
\def\mCAfe{$6\pm1$}
\def\mCDensityOne{$18.0^{+1.4}_{-0.2}$}
\def\mCDensityTwo{$6.2^{+19.5}_{-0.3}$}
\def\mCRfOne{$0.475^{+0.004}_{-0.047}$}
\def\mCRfTwo{$<6$}
\def\mCnrelOne{$14.89^{+0.54}_{-0.01}$}
\def\mCnrelTwo{$0.7^{+0.5}_{-0.4}$}
\def\mCLineE{$6.97\pm0.02$}
\def\mCSigma{$130^{+30}_{-40}$}
\def\mCStrength{$0.046\pm0.005$}
\def\mCNHxstar{$10.8^{+0.2}_{-3.1}$}
\def\mClogxixstar{$4.09^{+0.04}_{-0.04}$}
\def\mCvxstar{$2100^{+600}_{-600}$}
\def\mCNccFPMA{$1^a$}
\def\mCNccHEXTE{$0.918\pm0.001$}
\def\mCNccSwift{$0.859\pm0.004$}
\def\mCGauPCAoneN{$0.16\pm0.05$} 
\def\mCGauPCAtwoN{$0.09\pm0.05$} 
\def\mCGauSwiftoneN{$5\pm1$} 
\def\mCGauSwifttwoN{$4^{+2}_{-1}$} 
\def\mCGauSwiftoneSigma{$<0.07$} 
\def\mCGauSwifttwoSigma{$0.13\pm0.04$} 
\def\mCchi{$211$}
\def\mCnu{$188$}
\def\mCchired{$1.12$}

We model the full broadband, combined \rxte\ and \swift-XRT spectrum of \j1752 (as described in detail in Section~\ref{sec:data}) with a high-density reflection model, exploring different setups to investigate the interplay between the thermal disk blackbody and reflection components. In addition, we seek to test the effects of invoking a more complex coronal continuum, in lieu of the recent results of \citetalias{Zdziarski2021a}. We use {\it three} classes of model: A) high-density relativistic and distant reflection from a Comptonized disk; B) high-density relativistic and distant reflection from a Comptonized disk that is constrained by the $R_{\rm in}$ parameter of the reflection model; and C) two high-density reflection components from different coronal illumination spectra and a disk blackbody. In full component form the models are defined as follows:

\begin{itemize}
\item Model A: {\tt crabcorr*TBabs(simplcut$\otimes$ezdiskbb + relxillDCp + xillverDCp + Gau + Gau + Gau + Gau)}, $N_{\rm disk}$ free, $\log n_{\rm e}$ free
\item Model A.2: {\tt crabcorr*TBabs(simplcut$\otimes$ezdiskbb + relxillDCp + xillverDCp + Gau + Gau + Gau + Gau)}, $N_{\rm disk}$ free, $\log n_{\rm e} = 20$
\item Model B: {\tt crabcorr*TBabs(simplcut$\otimes$ezdiskbb + relxillDCp + xillverDCp + Gau + Gau + Gau + Gau)}, $N_{\rm disk}$ tied to $R_{\rm in}$, $\log n_{\rm e}$ free
\item Model C: {\tt crabcorr*TBabs(ezdiskbb + relxillDCp + relxillDCp2 + Gau + Gau + Gau + Gau)}, i.e., double corona and reflection, $N_{\rm disk}$ free, $\log n_{\rm e}$ free
\end{itemize}

The four additional Gaussian components represent systematic emission features in both the \swift-XRT and \rxte-PCA-and-HEXTE detectors respectively. As discussed in detail in Section~\ref{sec:swift_modeling}, the \swift-XRT spectrum contains residual features around $\sim2$~keV. In the combined spectrum we can identify two emission features at $\sim1.7$~keV and $\sim2.3$~keV respectively, which we model with Gaussian lines fixed at those energies with variable width, $\sigma$. The additional two Gaussian lines are fixed at $\sim29.8$~keV and $43.2$~keV respectively, with their width fixed to $\sigma=0.1$---see Section~\ref{subsubsec:ld-hard} for a full description. We also refer the reader to both \citetalias{Garcia2018a} and \citetalias{Zdziarski2021a} for further description of the details of these features. The {\tt relxillDCp} and {\tt xillverDCp} components are the same as the frequently adopted {\tt relxillCp} and {\tt xillverCp} flavours of {\tt relxill} \citep{Garcia2014,Dauser2014}, except with the additional free density parameter, ranging from $n_{\rm e}=10^{15}\mbox{--}10^{20}~{\rm cm^{-3}}$ \citep{Garcia2016}.


\begin{deluxetable*}{llcccc}
\tabletypesize{\footnotesize}
\tablecaption{Maximum likelihood estimates of all parameters in spectral fitting of broadband, combined \rxte\ and \swift-XRT observations of the hard state plateau of \j1752. \label{tab:plateau-params}}
\tablecolumns{6}
\tablehead{
\colhead{Component} &
\colhead{Parameter} & 
\colhead{Model A} &
\colhead{Model A2} & 
\colhead{Model B} & 
\colhead{Model C} \\
}
\startdata
{\tt crabcorr} & $N_{\rm CC, PCA}$ & \multicolumn{4}{c}{$1.097^a$}  \\
{\tt crabcorr} & $\Delta\Gamma_{\rm CC, PCA/HEXTE}$ & \multicolumn{4}{c}{$0.01^a$} \\
{\tt crabcorr} & $\Delta\Gamma_{\rm CC, XRT}$ & \multicolumn{4}{c}{$-0.04^a$} \\
{\tt relxillDCp} & $a_{\star}$ & \multicolumn{4}{c}{$0.998^a$}  \\
{\tt TBabs} & $N_{\rm H}~[10^{22}~{\rm cm^{-2}}]$ & \multicolumn{4}{c}{$1^a$} \\
{\tt Gaussian} & $E_{\rm g1}~{\rm [keV]}$ & \multicolumn{4}{c}{$1.7^a$} \\
{\tt Gaussian} & $E_{\rm g2}~{\rm [keV]}$ & \multicolumn{4}{c}{$2.3^a$} \\
{\tt Gaussian} & $E_{\rm g3}~{\rm [keV]}$ & \multicolumn{4}{c}{$29.8^a$} \\
{\tt Gaussian} & $E_{\rm g4}~{\rm [keV]}$ & \multicolumn{4}{c}{$43.2^a$} \\
\hline
{\tt simplcut/relxillDCp} & $\Gamma$ & \mAga\ & \mATwoga\ & \mBga\  &  \mCgaOne\  \\
{\tt simplcut/relxillDCp} & $kT_{\rm e}$~[keV] & \mAkTe\ & \mATwokTe\ & \mBkTe\ & \mCkTeOne\  \\
{\tt simplcut} & $f_{\rm sc}$ & \mAfsc\ & \mATwofsc\ & \mBfsc  & \nodata\  \\
{\tt ezdiskbb} & $kT_{\rm in}$~[keV] & \mATin & \mATwoTin\ & \mBTin\  & \mCTin\  \\
{\tt ezdiskbb} & $N_{\rm disk}~{\rm [10^3]}$ & \mAndisk & \mATwondisk\ & \mBndisk\  & \mCndisk\  \\
{\tt ezdiskbb} & $\Sigma$ & \nodata\ & \nodata\ & \mBSIGMA\  & \nodata\  \\
{\tt relxillDCp} & $q$ & \mAq\ & \mATwoq\ & \mBq\ & \mCq\  \\
{\tt relxillDCp} & $\log{\xi}$~[${\rm erg~cm~s^{-1}}$] & \mAlogxi & \mATwologxi\ &  \mBlogxi\  &  \mClogxiOne\  \\
{\tt relxillDCp} & $A_{\rm Fe}$~[Solar] & \mAAfe\ & \mATwoAfe\ & \mBAfe\  & \mCAfe\ \\
{\tt relxillDCp} & $i$~[$^\circ$] & \mAincl\ & \mATwoincl\ & \mBincl\  & \mCincl\  \\
{\tt relxillDCp} & $R_{\rm in}~[R_{\rm ISCO}]$ & \mARin\ & \mATwoRin\ & \mBRin\ & \mCRinOne\  \\
{\tt relxillDCp} & $\log n_{\rm e}~{\rm [cm^{-3}]}$ & \mADensityOne\ & \mATwoDensityOne\ & \mBDensityOne\  & \mCDensityOne\  \\
{\tt relxillDCp} & $R_{\rm f}$ & \nodata\ & \nodata\ & \nodata\ & \mCRfOne\  \\
{\tt relxillDCp} & $N_{\rm rel}~[10^{-3}]$ & \mAnrelOne\ & \mATwonrelOne\ & \mBnrelOne\   & \mCnrelOne\  \\
{\tt xillverDCp} & $N_{\rm xil}~[10^{-3}]$ & \mAnrelTwo\ & \mATwonrelTwo\ & \mBnrelTwo\   & \nodata\  \\
{\tt relxillDCp2} & $\Gamma_2$ & \nodata\ & \nodata\ & \nodata\ & \mCgaTwo\  \\
{\tt relxillDCp2} & $kT_{\rm e, 2}$~[keV] & \nodata\ & \nodata\ & \nodata\ & \mCkTeTwo\  \\
{\tt relxillDCp2} & $\log{\xi_2}$~[${\rm erg~cm~s^{-1}}$] & \nodata\ & \nodata\ &  \nodata\  &  \mClogxiTwo\  \\
{\tt relxillDCp2} & $R_{\rm in, 2}~[R_{\rm ISCO}]$ & \nodata\ & \nodata\ & \nodata\ & \mCRinTwo\  \\
{\tt relxillDCp2} & $R_{\rm f, 2}$ & \nodata\ & \nodata\ & \nodata\ & \mCRfTwo\  \\
{\tt relxillDCp2} & $N_{\rm rel, 2}~[10^{-3}]$ & \nodata\ & \nodata\ & \nodata\ & \mCnrelTwo\  \\
{\tt Gaussian} & $N_{\rm g1}~[10^{-3}]$ & \mAGauSwiftoneN & \mATwoGauSwiftoneN & \mBGauSwiftoneN & \mCGauSwiftoneN \\
{\tt Gaussian} & $\sigma_{\rm g1}~{\rm [keV]}$ & \mAGauSwiftoneSigma & \mATwoGauSwiftoneSigma & \mBGauSwiftoneSigma & \mCGauSwiftoneSigma \\
{\tt Gaussian} & $N_{\rm g2}~[10^{-3}]$ & \mAGauSwifttwoN & \mATwoGauSwifttwoN & \mBGauSwifttwoN & \mCGauSwifttwoN \\
{\tt Gaussian} & $\sigma_{\rm g2}~{\rm [keV]}$ & \mAGauSwifttwoSigma & \mATwoGauSwifttwoSigma & \mBGauSwifttwoSigma & \mCGauSwifttwoSigma \\
{\tt Gaussian} & $N_{\rm g3}~[10^{-3}]$ & \mAGauPCAoneN & \mATwoGauPCAoneN & \mBGauPCAoneN & \mCGauPCAoneN \\
{\tt Gaussian} & $N_{\rm g4}~[10^{-3}]$ & \mAGauPCAtwoN & \mATwoGauPCAtwoN & \mBGauPCAtwoN & \mCGauPCAtwoN \\
{\tt crabcorr} & $N_{\rm CC, HEXTE}$ & \mANccHEXTE\ & \mATwoNccHEXTE\ & \mBNccHEXTE\  & \mCNccHEXTE\  \\
{\tt crabcorr} & $N_{\rm CC, \swift}$ & \mANccSwift\ & \mATwoNccSwift\ & \mBNccSwift\ & \mCNccSwift\  \\
\hline
&  $\chi^2$ & \mAchi\ & \mATwochi\ &  \mBchi\ & \mCchi\  \\
&  $\nu$ & \mAnu\ &  \mATwonu\ & \mBnu\ & \mCnu\  \\
& $\chi_{\nu}^2$ & \mAchired\ &  \mATwochired\ & \mBchired\ & \mCchired\  \\
\enddata
\tablecomments{Model A: {\tt crabcorr*TBabs(simplcut$\otimes$ezdiskbb + relxillDCp+xillverCp+Gau+Gau+Gau+Gau)} (Model A.2 is equivalent in formulation, but with $\log n_{\rm e}=20$). Model B: {\tt crabcorr*TBabs(simplcut$\otimes$ezdiskbb+relxillDCp+xillverDCp+Gau+Gau+Gau+Gau)}. Model C: {\tt crabcorr*TBabs(ezdiskbb+relxillDCp+relxillDCp2+Gau+Gau+Gau+Gau)}. $N_{\rm CC}$ and $\Delta\Gamma_{\rm CC}$ are the normalization and photon index shifts in the component {\tt crabcorr}, shown in the table for each instrument. The disk normalization is given by $N_{\rm disk} = (R_{\rm in}/\kappa^2 D_{10})^2\cos i$, where $R_{\rm in}$ is the apparent inner disk in km, $D_{10}$ is the distance to the source in units of 10~kpc, $i$ is the disk inclination, and $\kappa$ is the color correction factor. The total $\chi^2$ is shown for each fit, along with the degrees of freedom, $\nu$, and the reduced $\chi^2$, $\chi^2_{\nu}=\chi^2/\nu$. The ionization, $\log\xi$, is given by $4\pi F_{\rm irr}/n_{\rm e}$, where $F_{\rm irr}$ is the ionizing flux, and $n_{\rm e}$ is the gas density. The normalization definition of the reflection models, given by $N_{\rm rel}$ and $N_{\rm xil}$, is such that the integrated energy flux from $0.1\mbox{--}1000$~keV is equal to $10^{20}n_{\rm e}\xi/4\pi$. }
\tablenotetext{a} {Frozen parameter}
\label{tab:fits}
\end{deluxetable*}


The disk and Comptonization spectral parameters are free to vary: disk normalization $N_{\rm disk}$, peak disk temperature $T_{\rm in}$, photon index $\Gamma$, electron temperature $kT_{\rm e}$, and scattering fraction $f_{\rm sc}$ (the fraction of photons that are Compton scattered). In the reflection components of Models A and B, we fix the black hole spin to maximal, $a=0.998$, and reflection fraction $R_{\rm f}=-1$ such that we return only the reflected component, allowing the normalization, $N_{\rm rel}$, to vary freely. The free parameters of {\tt relxillDCp} are the emissivity index, $q$, disk inclination, $i$, disk inner radius, $R_{\rm in}$, ionization, $\log\xi$, iron abundance, $A_{\rm Fe}$, and density, $n_{\rm e}$. The {\tt xillverDCp} parameters are tied to those of {\tt relxillDCp} with the exception of the ionization, $\log\xi$, which is frozen to 0, giving an ionization of 1, or neutral. This includes the density parameter, $n_{\rm e}$, which is based upon our ignorance regarding the radial density profile, since it is generally a highly model-dependent estimate, using both analytical and numerical approaches \citep{Svensson1994,Liska2022}. The only freely varying parameter of {\tt xillverDCp} is its normalization, $N_{\rm xil}$. We also fix the hydrogen column density in {\tt TBabs}, $N_{\rm H}=10^{22}~{\rm cm^{-2}}$, based upon our preliminary fits shown in the previous Sections. 

Models A and B are essentially identical but for the treatment of the disk normalization parameter, $N_{\rm disk}$, which in the latter case is tied to the inner disk radius parameter of {\tt relxillDCp} through an approximate prescription of the connection between disk area and flux using Equation~\ref{eq:diskbb}. As described in Section~\ref{subsubsec:ld-hard} we account for the uncertainties in $M_{\rm BH}$, $i$, $D$, and $\kappa$, by introducing the variable $\Sigma\propto(M_{\rm BH}^2 \cos i / D^2\kappa^4)$. $\Sigma$ essentially acts as a rescaling of the baseline characteristics, set at $M_{\rm BH}=10~M_{\odot}$, $i=30^{\circ}$, $D=3.5~{\rm kpc}$, and $\kappa=1.7$. Therefore in the fitting procedure the disk normalization scales as $N_{\rm disk}\propto \Sigma R_{\rm in}^2$. In this way the intrinsic disk blackbody emission as described by {\tt ezdiskbb} is limited by the reflection component constraints on $R_{\rm in}$ due to the broadness of the iron line, predominantly, with a rescaling applied to account for uncertainties in the system characteristics. The Model flavor A.2 is equivalent to Model A except we fix the disk density to its maximal value in the {\tt relxillDCp} tables---$10^{20}~{\rm cm^{-3}}$. 

Model C differs quite significantly from Models A and B in that we now adopt two separate coronal and associated reflection components, allowing more freedom in the aggregate shape of the underlying continuum, as well as in the total reflection spectrum. Here we allow the same freedom in the principle reflection component, {\tt relxillDCp}, as allowed in Models A and B, but then untie the coronal continuum parameters of the second component, {\tt relxillDCp2}, from the first, such that each continuum has a separate $\Gamma$ and $kT_{\rm e}$. We then freeze the ionization of the second component to $\log\xi=0$, its emissivity to $q=3$ (the non-relativistic limit), and tie the iron abundance, disk inclination, and disk density, to those of the first component. In a sense, the secondary reflection component acts as a proxy for the distant reflector, but with the additional freedom of some level of relativistic broadening and full integration over the disk, as well as irradiation from a secondary coronal continuum.

Model C is a loose representation of the two-component model presented by \cite{Zdziarski2021b} and applied to the hard state spectrum of MAXI~J1820$+$020. The only differences are that: (i) here we do not invoke a hybrid distribution of hot electrons in the secondary corona, since there is no requirement for a high-energy tail in the HEXTE data; and (ii) we do not impose a strong preference for the locations of the hard and soft coronal components, i.e., we allow the model to vary arbitrarily in terms of which component illuminates the inner or outer disk regions. We set the reflection fractions ($R_{\rm f}$) to positive values (such that the reflection model includes both the irradiating coronal continuum and the reflected component) and allow them to vary freely, as well as the normalizations of both reflection components ($N_{\rm rel}$ and $N_{\rm rel, 2}$). 

\begin{figure*}
\centering
\includegraphics[width=0.8\linewidth]{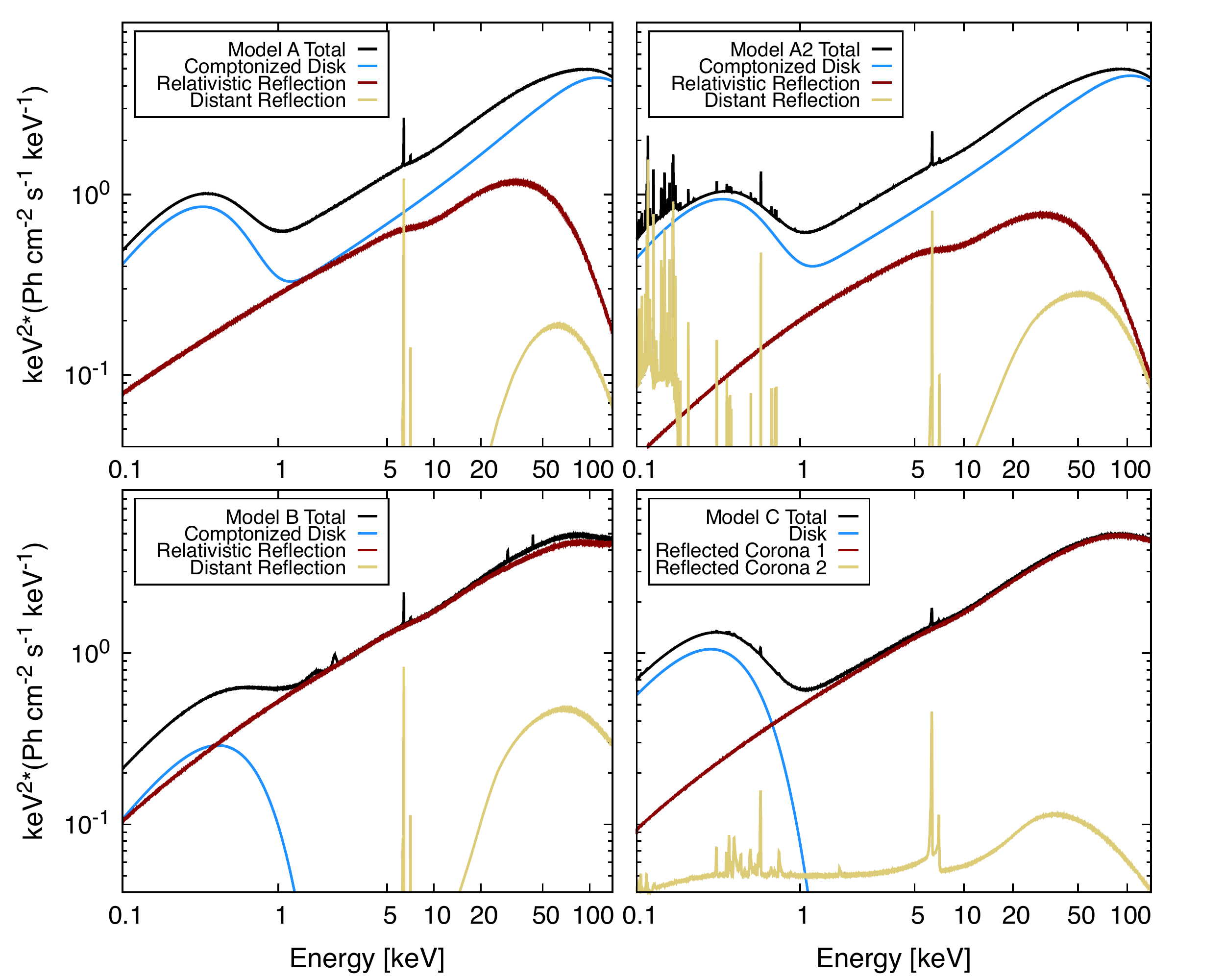}
\caption{The unabsorbed model components associated with each of the four models fit to the total combined \swift\ and \rxte\ spectrum (Models A, A2, B, and C). The total model is shown in black, Comptonized disk component in blue, relativistic reflection in red, and distant reflection in gold. In the bottom right panel (Model C), the distant reflector is actually a {\tt relxillDCp} component with a large truncation radius, $R_{\rm in}$.}
\label{fig:models}
\end{figure*}

Table~\ref{tab:fits} shows the best fit parameters and their associated uncertainties for fits of all models to the total combined \swift-XRT/\rxte\ spectra. There are key distinct differences between the fits, as well as stark contrasts with the results of both \citetalias{Garcia2018a} and \citetalias{Zdziarski2021a}. There are, however, several commonalities between the fits, particularly with regards to the reflection properties. We find high ionization ($\log\xi>4$) and iron abundance ($A_{\rm Fe}\ge5$), and minimal disk truncation $R_{\rm in}$ a few $R_{\rm ISCO}$, across all models. Models B and C have harder ($\Gamma\sim1.35$) coronal components than the Model A variants ($\Gamma\sim1.4\mbox{--}1.5$). The two coronal components in Model C appear to differ as expected (though we note $kT_{\rm e, 2}>100$~keV is merely a lower limit), with a harder Comptonization component ($\Gamma=1.36\pm0.02$, $kT_{\rm e}=97.7^{+5.4}_{-0.1}$), and a softer component.   

It is worth evaluating the implications of the fundamental difference in disk properties between Models A and B. In Model B, the disk normalization is constrained such that the area of the inner disk matches that implied by the inner disk radius ($R_{\rm in}$) constraint of the reflection component, and the uncertainties on $M_{\rm BH}$, $i$, $D$, and $\kappa$, via the $\Sigma$ parameter. The disk radius is strongly constrained by the PCA data in the form of both the red wing of the iron emission line, and the strength/shape of the smeared iron edge on the blue side of the line (as we showed in detail in Section~\ref{subsubsec:ld-hard}). This results in a very low disk normalization of course, $N_{\rm disk}=8800^{+1500}_{-900}$, roughly two orders of magnitude lower than found with the other model fits (A, A2, and C). This highlights the discrepancy between reflection and intrinsic disk emission constraints, which can be described as minor in spatial terms, but statistically significant. 



\begin{figure*}
\centering
\includegraphics[width=0.8\linewidth,trim={0 2cm 0 0}]{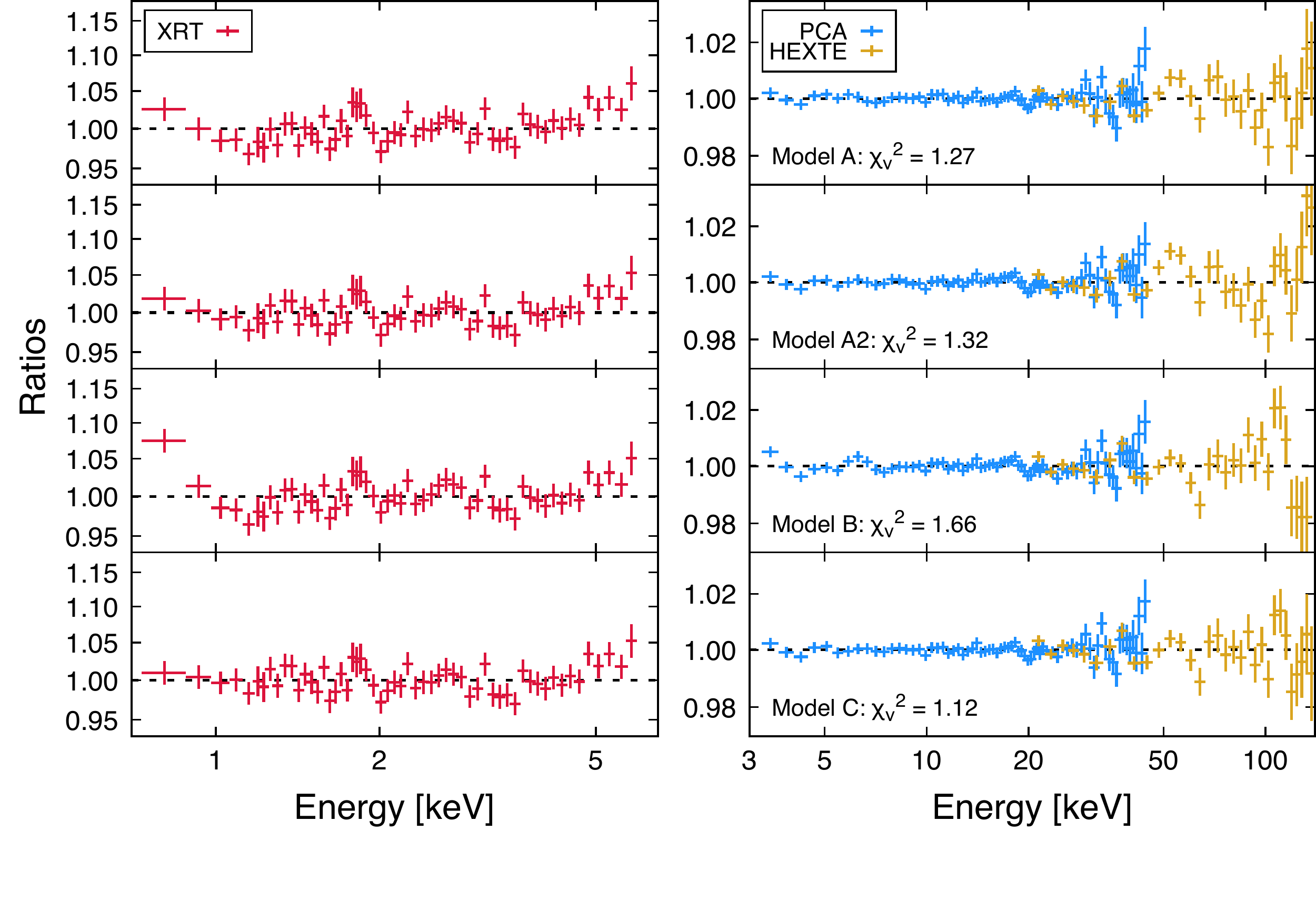}
\caption{Ratios of the final model fits to the combined \swift\ and \rxte\ spectrum (Models A, A2, B, and C). The \swift\ and \rxte\ data are shown in separate panels (left vs right) with variable y axis limits for clarity. Residuals remain around the $1.7\mbox{--}2$~keV region of the \swift-XRT spectrum due to the model Gaussians not perfectly modeling out the instrumental features in that range.}
\label{fig:final_ratios}
\end{figure*}

Figure~\ref{fig:models} shows a comparison of each best fit model. The high-density effects, in the form of partially thermalized remission, fail to fully subsume the direct disk emission, and the much fainter disk contributes to some of the soft excess emission. Figure~\ref{fig:final_ratios} shows the ratio residuals for all four model fits, divided between \swift-XRT (left) and \rxte\ (right) data for clarity. Model~B, for which the disk radius is constrained by the reflection model, displays residuals below 1~keV, showing that the model cannot entirely account for the soft emission. This confirms that there is a disagreement between the inner disk radius and the reflection radius. The additional concern with Model~B is that the scattering fraction, $f_{\rm sc}$ is very low, which can be understood by the lack of a visible powerlaw component (blue) in the bottom left panel of Figure~\ref{fig:models}. This implies a very high reflection fraction in order to make the model fit to the data. Given we also find a shallow emissivity profile in the Model~B fit ($q=3.74^{+0.50}_{-0.07}$; almost consistent with the Newtonian approximation), it is difficult to explain such a high reflection fraction.

\begin{figure}
\centering
\vspace{-1cm}
\includegraphics[width=0.98\linewidth]{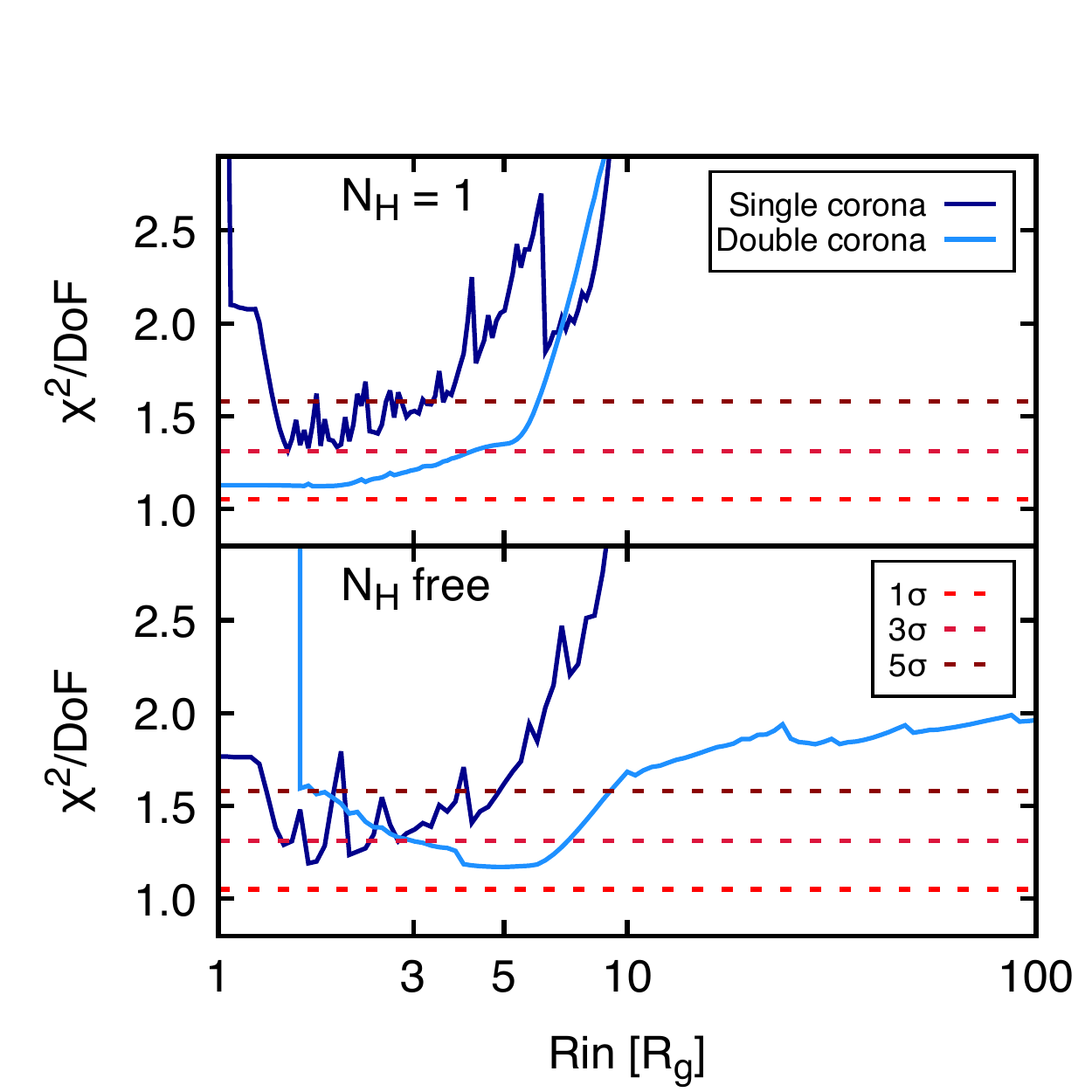}
\caption{The $\chi^2$ landscape of the inner disk radius parameter, $R_{\rm in}$, in units of $R_{\rm g}$, showing a comparison of constraints between applying Model A (single corona; dark blue) vs Model C (double corona; light blue) to the combined broadband \swift/\rxte\ observations of \j1752. $\chi^2$ has been calculated by running the Xspec routine {\tt steppar} in 200 $\log_{10}$ steps from $1\mbox{--}100~R_{\rm g}$. In the top panel the hydrogen column density is fixed at $N_{\rm H}=10^{22}~{\rm cm^{-2}}$ as in the final fits, whereas in the bottom panel we show the resulting contours with $N_{\rm H}$ free. The $1\sigma$, $3\sigma$, and $5\sigma$ limits on the $\chi^2$ distribution are shown to emphasize the constraints.}
\label{fig:rin_contours}
\end{figure}

Figure~\ref{fig:rin_contours} shows a comparison between the $R_{\rm in}$ fit landscape in Models A and C, showing the impact of invoking a secondary coronal component on the resulting truncation constraint---this can be seen as an attempt to test the predictions made by \citetalias{Zdziarski2021a}. We choose to compare Models~A and C both because they provide the best fits to the data, and they are directly comparable ($n_{\rm e}$ and the $N_{\rm disk}$ are free to vary in both). We find that when allowing the additional freedom of a secondary IC continuum and associated reflection component, $R_{\rm in}$ is still constrained by the data to be less than $10~R_{\rm g}$ at $5\sigma$ significance, even with a variable hydrogen column density ($N_{\rm H}$). We note that our modeling approach naturally allows broad---and arguably unrealistic---freedom regarding the combination of the two coronal components; the photon indices, $\Gamma$, and coronal temperatures, $kT_{\rm e}$ are independently varying over a wide range of possible values, and whilst we constrain the ionization and emissivity of the secondary reflection component, there is significant freedom provided by variable inner disk radius in both components. Despite this freedom to the total model, the primary reflection component still requires that the disk be only mildly truncated, increasing by a factor of $\sim3$ with respect to a single Comptonization/reflection model. 


We can verify the validity of our geometrical constraints and the connection between $R_{\rm in}$, $n_{\rm e}$, and the irradiating flux, by calculating a rough estimate of the ionization of the inner accretion disk, given by $\xi=L_{\rm irr}/n_{\rm e} R_{\rm in}^2$, where $L_{\rm irr}$ is the irradiative (coronal) luminosity. Taking the approximate observed model coronal flux in the $0.1\mbox{--}1000$~keV band to be the total observed flux, $2\times10^{-8}~{\rm erg~cm^{-2}~s^{-1}}$, and assuming $D\sim3.5$~kpc and $M_{\rm BH}=10~M_{\odot}$, we find $L_{\rm IC}\sim2.8\times10^{37}~{\rm erg~s^{-1}}$, or $\sim0.02~L_{\rm Edd}$. If we adopt the maximal disk density of the reflection model, consistent with the application of Model~A2, $n_{\rm e}=10^{20}~{\rm cm}^{-3}$, and assume the disk is truncated at $R_{\rm in}=2~R_{\rm g}$ (as found for Model~A2 fits), we find $\log\xi \sim 4.5~{\rm erg~s^{-1}~cm}$. This value is a little higher than that found in fits of Model~A2 ($\log\xi=4.32^{+0.01}_{-0.02}~{\rm erg~s^{-1}~cm}$), but it is roughly consistent.

\section{Discussion and Conclusions}\label{sec:discussion}

We have conducted an in depth analysis and modeling of the long-stable hard state of \j1752\ using combined simultaneous \swift-XRT and \rxte\ observations over a 10-day and month-long period respectively. 

We have shown, unequivocally, that the Fe K line and edge observed during the hard state are broad, and these broad components cannot be subsumed by invoking an additional coronal component in the underlying continuum. Another way to frame the problem is that there may be multiple coronal IC components present, thus making the irradiating continuum complex, but this does not replace the need for broad reflection features; the reflecting material must still be within $10~R_{\rm ISCO}$ at the $5\sigma$ significance level. 

We have also shown that the intrinsic disk emitting flux implies only a mild disagreement between the ``thermal radiation" and ``reflection" edges of the inner disk. The value of $R_{\rm in}$ derived from the disk normalization constraint is consistently larger than that constrained by the reflection model, but only by a factor of a few, and at the statistical extremes they are remarkably close ($R_{\rm in}<5~R_{\rm ISCO}$ from reflection, $R_{\rm in}>6~R_{\rm ISCO}$ from disk emission modeling, assuming limits of $D=3~{\rm kpc}$, $M_{\rm BH}=11~M_{\odot}$, $i=10^{\circ}$ and $\kappa=1.2$). It is worth exploring these two key findings in more detail, before digging deeper into the implications of our other parameter constraints, and the comparisons with both \citetalias{Garcia2018a} and \citetalias{Zdziarski2021a}. 

\subsection{The multiple-Comptonization hypothesis and reflection}
\label{subsec:multi_corona}

There exist several arguments in favor of both multiple coronal emission components and high disk truncation in \j1752 during its hard state. Here we address some of these arguments in the context of our own spectral fitting results. 

Firstly, that the power spectrum of \j1752, as with many hard state BHBs, contains multiple power spectral components (represented as broad Lorentzians in frequency space, see Figure~3 of \citetalias{Zdziarski2021a}). The rationale here is that it is difficult to explain this level of complexity in the X-ray variability of the source if the corona is simply a compact, lamppost-like region illuminating the disk; it implies at least some coronal extension, and perhaps independently varying coronae. This is a sensible argument, and one to consider seriously, but as we have shown, it does not necessarily resolve the question of how truncated the thin accretion disk is in the hard state. 

Secondly, \citetalias{Zdziarski2021a} claim that spectral fits that assume a single-Comtponization continuum model violate pair equilibrium balance, i.e., the compactness of the corona implies pair runaway. \citetalias{Zdziarski2021a} reach this conclusion based upon the coronal lamppost height, $h\sim1.7~R_{\rm g}$, and coronal temperature, $kT_{\rm e}=174\pm17$~keV, constrained by their spectral modeling. The coronal compactness parameter is given by $l=L_{\rm IC} \sigma_{\rm T}/Rm_{\rm e}c^3\sim2.1\times10^4$. Using Figure~1 of \cite{Fabian2015} as a benchmark of the $l-\theta_{\rm e}$ (compactness-temperature) plane, \citetalias{Zdziarski2021a} argue that for the temperatures they find in their spectral fits, the ${\rm e}^{\pm}$ runaway limit is at $l\sim10$. Clearly the derived compactness for the given geometry, luminosity, and coronal temperature, violates this limit. Therefore it seems single-Comptonization coronal continua can lead to unphysical results, given the necessity for high irradiative emissivities to achieve the desired fits to the reflection component. 

We find, however, significantly lower coronal electron temperatures in our single-Comptonization fits, with the exception of Model~B (in which the disk spectrum is constrained by the reflection component inner radius). Models~A and A2 both yield $\sim50$~keV (see Table~\ref{tab:fits}), or $\theta_{\rm e}\sim0.1$. At these temperatures, compactnesses of $l>1000$ satisfy the ${\rm e}^{\pm}$ pair equilibrium assumption. Models~A and A2 adopt the {\tt relxillDCp} flavor of {\tt relxill}, as opposed to the lamppost geometry variants, so we cannot directly map the results to a coronal size. We can however make an estimate based upon the very high emissivity index ($q>9.5$, implying $h\sim1\mbox{--}3~R_{\rm g}$ based upon Figure~3 of \citealt{Dauser2013}). If we assume $R\sim2~R_{\rm g}$ for the coronal size, and take $\theta\sim0.1$ and the full $0.1\mbox{--}1000$~keV coronal luminosity $L_{\rm IC}\sim2\times10^{-8}~{\rm erg~cm^{-2}~s^{-1}}$, we find $l\sim250$. Upon inspection of Figure~1 of \cite{Fabian2015}, we see that our estimate does not violate the ${\rm e}^{\pm}$ runaway limit. We find a higher value in the case of Model~B, given the higher electron temperature $kT_{\rm e}\sim130$~keV, but in principle we can reject Model~B on more immediate grounds---that it does not fit the data as well as the other models. In short, it is not true that single-Comptonization model solutions universally violate the pair limit, it instead depends on the particulars of the spectral fitting parameter constraints. 

Another class of arguments relates specifically to the reflection component itself for the given disk-corona geometry. \citetalias{Zdziarski2021a} derive a lower limit on the inner disk radius, $R_{\rm in}$, based upon soft X-ray constraints on the remitted thermalized portion of the reprocessed photons. The arguments are based upon an analytical derivation provided by \cite{Zdziarski2020a}. In simplistic terms, the estimate is based upon the Stefan-Boltzmann law as applied to the thermalized disk, whereby the inner disk temperature is a quadrature summation of the intrinsic dissipation and irradiative heating: $T_{\rm in}^4=T_{\rm in, irr}^4+T_{\rm in, diss}^4$. \citetalias{Zdziarski2021a} give an estimate of the irradiative component as

\begin{equation}
\label{eq:zdz}
\sigma\left(\frac{T_{\rm in, irr}}{k}\right)^4 \approx \frac{(1-a)L_{\rm irr}}{2\pi R_{\rm in}^2} \,\,\, ,
\end{equation}

where $\sigma$ is the Stefan-Boltzmann constant, $k$ is the Boltzmann constant, and $a$ is the disk albedo. \citetalias{Zdziarski2021a} progress from the simple formalism presented in Equation~\ref{eq:zdz} to derive a lower limit of $R_{\rm in}\geq20~R_{\rm g}$. We stress, however, that in addition to the inherent uncertainties on the system characteristics, this estimate contains a large uncertainty in the approximation of the disk albedo, $a$, as well as an oversimplification of the process of thermalization. The disk albedo is likely both radial and energy dependent, as has been shown explicitly by, e.g., \cite{Kinch2019} and \cite{Taverna2020}. In fact, \cite{Taverna2020} show the dependency of $a$ on disk radius for various BH spins and photon energies, at high disk densities ($n_{\rm e}\sim10^{20}~{\rm cm^{-3}}$)---we note that their albedo profiles peak at high values ($a\sim0.6\mbox{--}1$) in the inner $5~r_{\rm g}$ of the disk. Whilst it is possible that the absolute irradiating flux and associated disk density should lead to stronger thermal emission than is accounted for by the modeling we present in this paper, this is by no means a hard limit on the inner radius due to the associated uncertainties with Equation~\ref{eq:zdz}. 

One can approach this bound on irradiative heating, whilst making fewer assumptions, by estimating the expected inner disk ionization, as we showed in Section~\ref{subsubsec:hd-hard}. This calculation tells us that the heating provided to the inner disk is at least consistent from the point of view of maintaining sufficient optical depth in iron to produce reflection features. 


Nonetheless, \citetalias{Zdziarski2021a} conclude that the single-Comptonization model with a non-truncated disk violates all these physical limits, and therefore a more complex continuum model with high disk truncation should be considered to solve this problem. In order to show the validity of this assumption, they attempt to fit the $\sim4\mbox{--}10$~keV PCA band with reflection from a truncated disk, finding $R_{\rm in}\geq90~R_{\rm g}$. We showed in Section~\ref{subsec:combined}/Figure~\ref{fig:hard_state_refl_ratios} that one cannot fit the broadband continuum of \j1752 with a reflection model that assumes high disk truncation. Furthermore, in fits of Model~C to the full spectral energy range ($0.55\mbox{--}140$~keV), we showed that even invoking a secondary coronal Comptonization component does not change the situation. Limiting the spectral data to the $4\mbox{--}10$~keV vastly reduces the robustness of reflection modeling constraints, which is clear from the sub-unity $\chi_{\nu}^2$ statistics found by \citetalias{Zdziarski2021a} in their attempts to model the $4\mbox{--}10$~keV spectrum\footnote{\citetalias{Zdziarski2021a} fit their reflection model, {\tt reflkerr}, with 5 free parameters, to 15 PCA channels, and find $\chi^2/{\rm DoF}=6/10$, as shown in their Table~3.}. 

We do however agree that the corona may be better represented as a multi-faceted region, as opposed to a single-temperature thermal plasma, based on some of the above arguments, in that it is a more physically-motivated proposition. Indeed adopting a multi-temperature coronal continuum improves our spectral fits, but it does not lead to high disk truncation ($R_{\rm in}$ remains low,  $\ll10~R_{\rm ISCO}$). 

\subsection{The discrepancy between the thermal and reflection disk ``edges"}
\label{subsec:disk_constraint}
An additional argument in favor of disk truncation in the hard state of \j1752 is that the flux in the observed disk emission component requires a truncated disk, regardless of the reflection model constraints (for example \citetalias{Zdziarski2021a} find $R_{\rm in}\approx100^{+100}_{-40}~R_{\rm g}$). In Section~\ref{subsubsec:hd-hard} we found that $R_{\rm in}=6\mbox{--}80~R_{\rm ISCO}$ at $90\%$ confidence as predicted by the disk constraint, accounting for uncertainties on the system characteristics. Our lower bound is smaller than found by \citetalias{Zdziarski2021a} due to the choice of disk model ({\tt ezdiskbb}), and likely the variable density allowed within the reflection component ({\tt relxillDCp}); higher density models see an excess of soft emission which can partially subsume the intrinsic disk emission component. We note however that this approach is not fully consistent in physical terms, as was pointed out by \citetalias{Zdziarski2021a}. The Stefan-Boltzmann law governs the thermal temperature of the disk effective surface, which is the sum of intrinsic and irradiative components, such that applying independent models for each (in the form of {\tt ezdiskbb} and {\tt relxillDCp}) is physically inconsistent. The intrinsic disk emission is not currently included in the {\tt xillver} reflection tables, but this will be a focus of future work, wherein we will be able to test a more consistent model. 

We also found that extending the reflection models to higher disk densities had no impact on the $R_{\rm in}$ constraint (as constrained by the reflection component) in modeling of this source. We see only a slight increase in $R_{\rm in}$ when adopting a multiple-Comptonization (two components) continuum model, roughly doubling its value to $\sim4~R_{\rm ISCO}$. As discussed in the previous Section, even with this increase we find at least a factor of $3$ disagreement between this reflection constraint and the intrinsic disk flux constraint.

There are often unconsidered nuances associated with the nature of the inner disk radius as a boundary. \cite{Krolik2002} explore in detail the characteristics ``edges" of an accretion disk: the ``turbulence edge", the ``stress edge", the ``reflection edge", and the ``radiation edge". We refer the reader to \cite{Krolik2002} for a full description of each, for now we choose to focus on a brief discussion of the latter two, the ``reflection edge" and ``radiation edge", since these in principle represent what we are able to measure with the reflection and disk continuum components in X-ray spectral fitting. Whilst previous works have typically addressed this question from the point of view of an accretion flow that already extends down to the ISCO (see, e.g., \citealt{Reynolds1997b, Young1998, Reynolds2008}), we can also view this problem from the standpoint of a truncated disk. {\it Could it be that in the bright hard states of BHBs, the thermal ``radiation edge" of the disk, the innermost location from which the blackbody emission originates, is marginally further from the ISCO than the ``reflection edge", the location from which the bulk of the reflection/reprocessing occurs?}

Recent radiation-transport two-temperature general relativistic magnetohydrodynamic (GRMHD) simulations have hinted at such a scenario \citep{Liska2022}, in which cold, optically thick clumps of gas may be infalling within a sharp magnetic-pressure-imposed disk truncation radius. \cite{Liska2022} suggest that these cold clumps of gas could explain the evidence for strong relativistic reflection close to the ISCO, without the need for a thin disk extending down to the ISCO. This may naturally explain the disconnect---which whilst statistically significant, is minor in our analysis---between the $R_{\rm in}$ constraint from reflection fitting, versus that from modeling of the intrinsic blackbody disk emission.

\subsection{Other parameter constraints and comparisons}
Another model parameter that stands out in our analysis is the iron abundance, $A_{\rm Fe}$. Recent works involving high-density reflection modeling have shown hints of a reduction in $A_{\rm Fe}$ with respect to lower-density reflection model constraints \citep{Tomsick2018,JJiang2019,JJiang2019b}. Here however, we find that the density parameter has little effect on the $A_{\rm Fe}$ constraint. We note that the constraints of \cite{Tomsick2018} and \cite{JJiang2019,JJiang2019b} were derived from spectral fits to data from modern, higher-energy-resolution X-ray instruments, such as \nustar, as well as being analyses of different sources. This could in principle be an indication that the spectral energy resolution plays a role in isolating the connection between $n_{\rm e}$ and $A_{\rm Fe}$. Adjusting the iron abundance in the model mostly impacts the strength of the Fe K line and Compton hump, whereas the density emphasizes the soft X-ray emission, whereby the energies at which the soft excess appears varies with density. The latter process generally has the effect of softening the reflection spectrum, which typically skews the underlying Comptonization continuum to be harder (we do not actually see this effect in our modeling, see the comparison of Models A and A2 in Table~\ref{tab:fits}).

 In the case of \j1752, the \swift-XRT data do not show strong soft emission, and the bulk of the soft excess is accounted for by the intrinsic disk spectrum, not the high-density reflected emission. Thus the model is not moving into a parameter space in which the high-density effects have an appreciable impact on the line region, and any impact that may be present is likely washed out by the lack of energy resolution at $5\mbox{--}7$~keV. It therefore could be expected that we see no change in the iron abundance between the fits. Interestingly, we do see an impact on the iron abundance when invoking two Comptonization components (Model~C; $A_{\rm Fe}=6\pm1$). When we fix $A_{\rm Fe}=1$ in Model~C fits, we find fits of comparable statistical quality to those of Models~A and A2. Thus, whilst adding complexity to the underlying continuum does not have much impact on inner radius constraints, it does appear to relax the typically observed super-Solar iron abundances. We also see no appreciable impact on the inner disk radius constraint, finding $R_{\rm in}=3^{+2}_{-1}~R_{\rm ISCO}$, consistent with the value shown in Table~1, in which $A_{\rm Fe}$ is variable.
 
\subsection{Conclusions}
We have performed a detailed and complete analysis of high signal-to-noise, combined \swift-XRT/\rxte\ spectra from observations of the long-stable ($\sim1$~month duration) hard state plateau of \j1752, during which the X-ray luminosity was $L_{\rm X}\sim0.02\mbox{--}0.1~L_{\rm Edd}$. We find that the inner disk radius, as determined via reflection spectroscopy, is close to the ISCO ($R_{\rm in}\sim1\mbox{--}5~R_{\rm ISCO}$; which we can otherwise view as the ``reflection edge" of the disk). The observed broad reflection features (the Fe K line and edge) cannot be explained purely as complexity in the underlying continuum with associated narrow reflection features. We also find that the soft flux, modeled by the intrinsic thermal disk blackbody emission, implies a certain amount of disk truncation ($R_{\rm in} = 6\mbox{--}80~R_{\rm ISCO}$), and find no physical limits that invalidate the low radius values. We note that whilst these two radius constraints differ, at their respective $90\%$ statistical upper/lower limits, they almost agree, with the latter intrinsic disk emission constraint being much less stringent than found with reflection modeling. 

We have shown that none of our solutions break the physical limits discussed by \citetalias{Zdziarski2021a} with regards to single-Comptonization reflection modeling: (i) pair equilibrium is satisfied; (ii) the inner disk ionization is roughly consistent with predictions from flux/density scalings, $\xi=L/n_{\rm e}R_{\rm in}^2$; (iii) The mismatch between the $R_{\rm in}$ constraints derived from the intrinsic disk emission and reflection modeling is not so dramatic (possibly a factor of $<2$), and we have discussed how this could arise naturally, with a sensible distinction between the ``reflection" and ``thermal radiation" edges. 

We conclude that the invocation of an additional coronal Comptonization component as the irradiating continua for reflection is worth exploring further, since we observe some impacts on the inner radius constraints from reflection modeling, as well as an impact on iron abundance estimates. We will apply similar modeling techniques---adopting high-density reflection models---to more sources in future works in order to study its impacts in a broader framework. 

 \acknowledgements
We thank the anonymous referee for their comments, which have served to improve this manuscript. 
 
 J.A.G. acknowledges support from NASA grant
NNX15AV31G and from the Alexander von Humboldt
Foundation. R.M.T.C. has been supported by NASA
grant 80NSSC177K0515. JJ acknowledges support from the Leverhulme Trust, the Isaac Newton Trust and St Edmund's College, University of Cambridge. JH acknowledges support from an appointment to the NASA Postdoctoral Program at the Goddard Space Flight Center, administered by the ORAU through a contract with NASA.
 
This research has made use of data, software and/or web tools obtained from the High Energy Astrophysics Science Archive Research Center (HEASARC), a service of the Astrophysics Science Division at NASA/GSFC and of the Smithsonian Astrophysical Observatory's High Energy Astrophysics Division. 

This research has made use of ISIS functions (ISISscripts) provided by 
ECAP/Remeis observatory and MIT (http://www.sternwarte.uni-erlangen.de/isis/). 

\vspace{5mm}
\facilities{\rxte\ (PCA; \citealt{Jahoda1996}, HEXTE; \citealt{Rothschild_1998a}), \swift-XRT \citep{Burrows2005}, HEASARC, HEASoft}

\software{{\tt XSPEC v.12.10.1s} \citep{Arnaud1996}, {\tt XILLVER} \citep{Garcia2010,Garcia2013}, {\tt RELXILL} (v1.3.11-5dev; \citealt{Garcia2014,Dauser2014}).}

\bibliographystyle{aasjournal}
\bibliography{references}

\begin{thebibliography}{}
\expandafter\ifx\csname natexlab\endcsname\relax\def\natexlab#1{#1}\fi
\providecommand{\url}[1]{\href{#1}{#1}}
\providecommand{\dodoi}[1]{doi:~\href{http://doi.org/#1}{\nolinkurl{#1}}}
\providecommand{\doeprint}[1]{\href{http://ascl.net/#1}{\nolinkurl{http://ascl.net/#1}}}
\providecommand{\doarXiv}[1]{\href{https://arxiv.org/abs/#1}{\nolinkurl{https://arxiv.org/abs/#1}}}

\bibitem[{{Anders} \& {Grevesse}(1989)}]{Anders1989}
{Anders}, E., \& {Grevesse}, N. 1989, \gca, 53, 197,
  \dodoi{10.1016/0016-7037(89)90286-X}

\bibitem[{{Arnaud}(1996)}]{Arnaud1996}
{Arnaud}, K.~A. 1996, in Astronomical Society of the Pacific Conference Series,
  Vol. 101, Astronomical Data Analysis Software and Systems V, ed. G.~H.
  {Jacoby} \& J.~{Barnes}, 17

\bibitem[{{Brocksopp} {et~al.}(2013){Brocksopp}, {Corbel}, {Tzioumis},
  {Broderick}, {Rodriguez}, {Yang}, {Fender}, \& {Paragi}}]{Brocksopp2013}
{Brocksopp}, C., {Corbel}, S., {Tzioumis}, A., {et~al.} 2013, \mnras, 432, 931,
  \dodoi{10.1093/mnras/stt493}

\bibitem[{{Brocksopp} {et~al.}(2010){Brocksopp}, {Corbel}, {Tzioumis},
  {Fender}, \& {Coriat}}]{Brocksopp2010}
{Brocksopp}, C., {Corbel}, S., {Tzioumis}, T., {Fender}, R., \& {Coriat}, M.
  2010, The Astronomer's Telegram, 2400, 1

\bibitem[{{Burrows} {et~al.}(2005){Burrows}, {Hill}, {Nousek}, {Kennea},
  {Wells}, {Osborne}, {Abbey}, {Beardmore}, {Mukerjee}, {Short}, {Chincarini},
  {Campana}, {Citterio}, {Moretti}, {Pagani}, {Tagliaferri}, {Giommi},
  {Capalbi}, {Tamburelli}, {Angelini}, {Cusumano}, {Br{\"a}uninger}, {Burkert},
  \& {Hartner}}]{Burrows2005}
{Burrows}, D.~N., {Hill}, J.~E., {Nousek}, J.~A., {et~al.} 2005, \ssr, 120,
  165, \dodoi{10.1007/s11214-005-5097-2}

\bibitem[{{Connors} {et~al.}(2017){Connors}, {Markoff}, {Nowak}, {Neilsen},
  {Ceccobello}, {Crumley}, {Froning}, {Gallo}, \& {Nip}}]{Connors2017}
{Connors}, R.~M.~T., {Markoff}, S., {Nowak}, M.~A., {et~al.} 2017, \mnras, 466,
  4121, \dodoi{10.1093/mnras/stw3150}

\bibitem[{{Connors} {et~al.}(2020){Connors}, {Garc{\'\i}a}, {Dauser},
  {Grinberg}, {Steiner}, {Sridhar}, {Wilms}, {Tomsick}, {Harrison}, \&
  {Licklederer}}]{Connors2020}
{Connors}, R. M.~T., {Garc{\'\i}a}, J.~A., {Dauser}, T., {et~al.} 2020, \apj,
  892, 47, \dodoi{10.3847/1538-4357/ab7afc}

\bibitem[{{Connors} {et~al.}(2021){Connors}, {Garc{\'\i}a}, {Tomsick}, {Hare},
  {Dauser}, {Grinberg}, {Steiner}, {Mastroserio}, {Sridhar}, {Fabian}, {Jiang},
  {Parker}, {Harrison}, \& {Kallman}}]{Connors2021}
{Connors}, R. M.~T., {Garc{\'\i}a}, J.~A., {Tomsick}, J., {et~al.} 2021, \apj,
  909, 146, \dodoi{10.3847/1538-4357/abdd2c}

\bibitem[{{Dauser} {et~al.}(2014){Dauser}, {Garc{\'{\i}}a}, {Parker}, {Fabian},
  \& {Wilms}}]{Dauser2014}
{Dauser}, T., {Garc{\'{\i}}a}, J., {Parker}, M.~L., {Fabian}, A.~C., \&
  {Wilms}, J. 2014, \mnras, 444, L100, \dodoi{10.1093/mnrasl/slu125}

\bibitem[{{Dauser} {et~al.}(2013){Dauser}, {Garcia}, {Wilms}, {B{\"o}ck},
  {Brenneman}, {Falanga}, {Fukumura}, \& {Reynolds}}]{Dauser2013}
{Dauser}, T., {Garcia}, J., {Wilms}, J., {et~al.} 2013, \mnras, 430, 1694,
  \dodoi{10.1093/mnras/sts710}

\bibitem[{{Fabian} {et~al.}(2015){Fabian}, {Lohfink}, {Kara}, {Parker},
  {Vasudevan}, \& {Reynolds}}]{Fabian2015}
{Fabian}, A.~C., {Lohfink}, A., {Kara}, E., {et~al.} 2015, \mnras, 451, 4375,
  \dodoi{10.1093/mnras/stv1218}

\bibitem[{{Fabian} {et~al.}(2014){Fabian}, {Parker}, {Wilkins}, {Miller},
  {Kara}, {Reynolds}, \& {Dauser}}]{Fabian2014}
{Fabian}, A.~C., {Parker}, M.~L., {Wilkins}, D.~R., {et~al.} 2014, \mnras, 439,
  2307, \dodoi{10.1093/mnras/stu045}

\bibitem[{{Garc{\'{\i}}a} {et~al.}(2013){Garc{\'{\i}}a}, {Dauser}, {Reynolds},
  {Kallman}, {McClintock}, {Wilms}, \& {Eikmann}}]{Garcia2013}
{Garc{\'{\i}}a}, J., {Dauser}, T., {Reynolds}, C.~S., {et~al.} 2013, \apj, 768,
  146, \dodoi{10.1088/0004-637X/768/2/146}

\bibitem[{{Garc{\'{\i}}a} \& {Kallman}(2010)}]{Garcia2010}
{Garc{\'{\i}}a}, J., \& {Kallman}, T.~R. 2010, \apj, 718, 695,
  \dodoi{10.1088/0004-637X/718/2/695}

\bibitem[{{Garc{\'{\i}}a} {et~al.}(2014){Garc{\'{\i}}a}, {Dauser}, {Lohfink},
  {Kallman}, {Steiner}, {McClintock}, {Brenneman}, {Wilms}, {Eikmann},
  {Reynolds}, \& {Tombesi}}]{Garcia2014}
{Garc{\'{\i}}a}, J., {Dauser}, T., {Lohfink}, A., {et~al.} 2014, \apj, 782, 76,
  \dodoi{10.1088/0004-637X/782/2/76}

\bibitem[{{Garc{\'\i}a} {et~al.}(2016){Garc{\'\i}a}, {Grinberg}, {Steiner},
  {McClintock}, {Pottschmidt}, \& {Rothschild}}]{Garcia2016}
{Garc{\'\i}a}, J.~A., {Grinberg}, V., {Steiner}, J.~F., {et~al.} 2016, \apj,
  819, 76, \dodoi{10.3847/0004-637X/819/1/76}

\bibitem[{{Garc{\'\i}a} {et~al.}(2018{\natexlab{a}}){Garc{\'\i}a}, {Kallman},
  {Bautista}, {Mendoza}, {Deprince}, {Palmeri}, \& {Quinet}}]{Garcia2018b}
{Garc{\'\i}a}, J.~A., {Kallman}, T.~R., {Bautista}, M., {et~al.}
  2018{\natexlab{a}}, in Astronomical Society of the Pacific Conference Series,
  Vol. 515, Workshop on Astrophysical Opacities, 282

\bibitem[{{Garc{\'\i}a} {et~al.}(2014){Garc{\'\i}a}, {McClintock}, {Steiner},
  {Remillard}, \& {Grinberg}}]{Garcia2014b}
{Garc{\'\i}a}, J.~A., {McClintock}, J.~E., {Steiner}, J.~F., {Remillard},
  R.~A., \& {Grinberg}, V. 2014, \apj, 794, 73,
  \dodoi{10.1088/0004-637X/794/1/73}

\bibitem[{{Garc{\'{\i}}a} {et~al.}(2015){Garc{\'{\i}}a}, {Steiner},
  {McClintock}, {Remillard}, {Grinberg}, \& {Dauser}}]{Garcia2015}
{Garc{\'{\i}}a}, J.~A., {Steiner}, J.~F., {McClintock}, J.~E., {et~al.} 2015,
  \apj, 813, 84, \dodoi{10.1088/0004-637X/813/2/84}

\bibitem[{{Garc{\'\i}a} {et~al.}(2018{\natexlab{b}}){Garc{\'\i}a}, {Steiner},
  {Grinberg}, {Dauser}, {Connors}, {McClintock}, {Remillard}, {Wilms},
  {Harrison}, \& {Tomsick}}]{Garcia2018a}
{Garc{\'\i}a}, J.~A., {Steiner}, J.~F., {Grinberg}, V., {et~al.}
  2018{\natexlab{b}}, \apj, 864, 25, \dodoi{10.3847/1538-4357/aad231}

\bibitem[{{Hill} {et~al.}(2004){Hill}, {Burrows}, {Nousek}, {Abbey}, {Ambrosi},
  {Br{\"a}uninger}, {Burkert}, {Campana}, {Cheruvu}, {Cusumano}, {Freyberg},
  {Hartner}, {Klar}, {Mangels}, {Moretti}, {Mori}, {Morris}, {Short},
  {Tagliaferri}, {Watson}, {Wood}, \& {Wells}}]{Hill2004}
{Hill}, J.~E., {Burrows}, D.~N., {Nousek}, J.~A., {et~al.} 2004, in Society of
  Photo-Optical Instrumentation Engineers (SPIE) Conference Series, Vol. 5165,
  X-Ray and Gamma-Ray Instrumentation for Astronomy XIII, ed. K.~A. {Flanagan}
  \& O.~H.~W. {Siegmund}, 217--231

\bibitem[{{Jahoda} {et~al.}(2006){Jahoda}, {Markwardt}, {Radeva}, {Rots},
  {Stark}, {Swank}, {Strohmayer}, \& {Zhang}}]{Jahoda_2006a}
{Jahoda}, K., {Markwardt}, C.~B., {Radeva}, Y., {et~al.} 2006, \apjs, 163, 401,
  \dodoi{10.1086/500659}

\bibitem[{{Jahoda} {et~al.}(1996){Jahoda}, {Swank}, {Giles}, {Stark},
  {Strohmayer}, {Zhang}, \& {Morgan}}]{Jahoda1996}
{Jahoda}, K., {Swank}, J.~H., {Giles}, A.~B., {et~al.} 1996, in Society of
  Photo-Optical Instrumentation Engineers (SPIE) Conference Series, Vol. 2808,
  EUV, X-Ray, and Gamma-Ray Instrumentation for Astronomy VII, ed. O.~H.
  {Siegmund} \& M.~A. {Gummin}, 59--70

\bibitem[{{Jiang} {et~al.}(2019{\natexlab{a}}){Jiang}, {Fabian}, {Wang},
  {Walton}, {Garc{\'\i}a}, {Parker}, {Steiner}, \& {Tomsick}}]{JJiang2019}
{Jiang}, J., {Fabian}, A.~C., {Wang}, J., {et~al.} 2019{\natexlab{a}}, \mnras,
  484, 1972, \dodoi{10.1093/mnras/stz095}

\bibitem[{{Jiang} {et~al.}(2019{\natexlab{b}}){Jiang}, {Fabian}, {Dauser},
  {Gallo}, {Garc{\'\i}a}, {Kara}, {Parker}, {Tomsick}, {Walton}, \&
  {Reynolds}}]{JJiang2019b}
{Jiang}, J., {Fabian}, A.~C., {Dauser}, T., {et~al.} 2019{\natexlab{b}},
  \mnras, 489, 3436, \dodoi{10.1093/mnras/stz2326}

\bibitem[{{Kinch} {et~al.}(2019){Kinch}, {Schnittman}, {Kallman}, \&
  {Krolik}}]{Kinch2019}
{Kinch}, B.~E., {Schnittman}, J.~D., {Kallman}, T.~R., \& {Krolik}, J.~H. 2019,
  \apj, 873, 71, \dodoi{10.3847/1538-4357/ab05d5}

\bibitem[{{Krolik} \& {Hawley}(2002)}]{Krolik2002}
{Krolik}, J.~H., \& {Hawley}, J.~F. 2002, \apj, 573, 754,
  \dodoi{10.1086/340760}

\bibitem[{{Kubota} {et~al.}(1998){Kubota}, {Tanaka}, {Makishima}, {Ueda},
  {Dotani}, {Inoue}, \& {Yamaoka}}]{Kubota1998}
{Kubota}, A., {Tanaka}, Y., {Makishima}, K., {et~al.} 1998, \pasj, 50, 667,
  \dodoi{10.1093/pasj/50.6.667}

\bibitem[{{Liska} {et~al.}(2022){Liska}, {Musoke}, {Tchekhovskoy}, {Porth}, \&
  {Beloborodov}}]{Liska2022}
{Liska}, M.~T.~P., {Musoke}, G., {Tchekhovskoy}, A., {Porth}, O., \&
  {Beloborodov}, A.~M. 2022, arXiv e-prints, arXiv:2201.03526.
\newblock \doarXiv{2201.03526}

\bibitem[{{Markoff} {et~al.}(2015){Markoff}, {Nowak}, {Gallo}, {Hynes},
  {Wilms}, {Plotkin}, {Maitra}, {Silva}, \& {Drappeau}}]{Markoff2015}
{Markoff}, S., {Nowak}, M.~A., {Gallo}, E., {et~al.} 2015, \apjl, 812, L25,
  \dodoi{10.1088/2041-8205/812/2/L25}

\bibitem[{{Markwardt} {et~al.}(2009){Markwardt}, {Swank}, {Barthelmy},
  {Baumgartner}, {Burrows}, {Evans}, {Holland }, {Hoversten}, \&
  {Page}}]{Markwardt2009b}
{Markwardt}, C.~B., {Swank}, J.~H., {Barthelmy}, S.~D., {et~al.} 2009, The
  Astronomer's Telegram, 2258, 1

\bibitem[{{Miller-Jones} {et~al.}(2011){Miller-Jones}, {Jonker}, {Ratti},
  {Torres}, {Brocksopp}, {Yang}, \& {Morrell}}]{Miller-Jones2011}
{Miller-Jones}, J.~C.~A., {Jonker}, P.~G., {Ratti}, E.~M., {et~al.} 2011,
  \mnras, 415, 306, \dodoi{10.1111/j.1365-2966.2011.18704.x}

\bibitem[{{Mitsuda} {et~al.}(1984){Mitsuda}, {Inoue}, {Koyama}, {Makishima},
  {Matsuoka}, {Ogawara}, {Shibazaki}, {Suzuki}, {Tanaka}, \&
  {Hirano}}]{Mitsuda1984}
{Mitsuda}, K., {Inoue}, H., {Koyama}, K., {et~al.} 1984, \pasj, 36, 741

\bibitem[{{Mu{\~n}oz-Darias} {et~al.}(2010){Mu{\~n}oz-Darias}, {Motta},
  {Pawar}, {Belloni}, {Campana}, \& {Bhattacharya}}]{MunozDarias2010}
{Mu{\~n}oz-Darias}, T., {Motta}, S., {Pawar}, D., {et~al.} 2010, \mnras, 404,
  L94, \dodoi{10.1111/j.1745-3933.2010.00842.x}

\bibitem[{{Nakahira} {et~al.}(2009){Nakahira}, {Negoro}, {Yamaoka}, {Matsuoka},
  {Ueda}, {Tomida}, {Sugizaki}, {Nakajima}, {Kawai}, {Yoshida}, \&
  {Mihara}}]{Nakahira2009}
{Nakahira}, S., {Negoro}, H., {Yamaoka}, K., {et~al.} 2009, The Astronomer's
  Telegram, 2259, 1

\bibitem[{{Nied{\'z}wiecki} {et~al.}(2019){Nied{\'z}wiecki}, {Szanecki}, \&
  {Zdziarski}}]{Niedzwiecki2019}
{Nied{\'z}wiecki}, A., {Szanecki}, M., \& {Zdziarski}, A.~A. 2019, \mnras, 485,
  2942, \dodoi{10.1093/mnras/stz487}

\bibitem[{{Novikov} \& {Thorne}(1973)}]{NT1973}
{Novikov}, I.~D., \& {Thorne}, K.~S. 1973, in Black Holes (Les Astres Occlus),
  343--450

\bibitem[{{Plotkin} {et~al.}(2012){Plotkin}, {Markoff}, {Kelly}, {K{\"o}rding},
  \& {Anderson}}]{Plotkin2012}
{Plotkin}, R.~M., {Markoff}, S., {Kelly}, B.~C., {K{\"o}rding}, E., \&
  {Anderson}, S.~F. 2012, \mnras, 419, 267,
  \dodoi{10.1111/j.1365-2966.2011.19689.x}

\bibitem[{{Reynolds}(1997)}]{Reynolds1997}
{Reynolds}, C.~S. 1997, \mnras, 286, 513, \dodoi{10.1093/mnras/286.3.513}

\bibitem[{{Reynolds} \& {Begelman}(1997)}]{Reynolds1997b}
{Reynolds}, C.~S., \& {Begelman}, M.~C. 1997, \apj, 488, 109,
  \dodoi{10.1086/304703}

\bibitem[{{Reynolds} \& {Fabian}(2008)}]{Reynolds2008}
{Reynolds}, C.~S., \& {Fabian}, A.~C. 2008, \apj, 675, 1048,
  \dodoi{10.1086/527344}

\bibitem[{{Rothschild} {et~al.}(1998){Rothschild}, {Blanco}, {Gruber},
  {Heindl}, {MacDonald}, {Marsden}, {Pelling}, {Wayne}, \&
  {Hink}}]{Rothschild_1998a}
{Rothschild}, R.~E., {Blanco}, P.~R., {Gruber}, D.~E., {et~al.} 1998, \apj,
  496, 538, \dodoi{10.1086/305377}

\bibitem[{{Shakura} \& {Sunyaev}(1973)}]{SS1973}
{Shakura}, N.~I., \& {Sunyaev}, R.~A. 1973, \aap, 24, 337

\bibitem[{{Shaposhnikov}(2010)}]{Shaposhnikov2010b}
{Shaposhnikov}, N. 2010, The Astronomer's Telegram, 2391, 1

\bibitem[{{Shaposhnikov} {et~al.}(2010){Shaposhnikov}, {Markwardt}, {Swank}, \&
  {Krimm}}]{Shaposhnikov2010a}
{Shaposhnikov}, N., {Markwardt}, C., {Swank}, J., \& {Krimm}, H. 2010, \apj,
  723, 1817, \dodoi{10.1088/0004-637X/723/2/1817}

\bibitem[{{Shaposhnikov} {et~al.}(2009){Shaposhnikov}, {Markwardt}, \&
  {Swank}}]{Shaposhnikov2009}
{Shaposhnikov}, N., {Markwardt}, C.~B., \& {Swank}, J.~H. 2009, The
  Astronomer's Telegram, 2269, 1

\bibitem[{{Steiner} {et~al.}(2017){Steiner}, {Garc{\'\i}a}, {Eikmann},
  {McClintock}, {Brenneman}, {Dauser}, \& {Fabian}}]{Steiner2017}
{Steiner}, J.~F., {Garc{\'\i}a}, J.~A., {Eikmann}, W., {et~al.} 2017, \apj,
  836, 119, \dodoi{10.3847/1538-4357/836/1/119}

\bibitem[{{Steiner} {et~al.}(2010){Steiner}, {McClintock}, {Remillard}, {Gou},
  {Yamada}, \& {Narayan}}]{Steiner2010}
{Steiner}, J.~F., {McClintock}, J.~E., {Remillard}, R.~A., {et~al.} 2010, \apj,
  718, L117, \dodoi{10.1088/2041-8205/718/2/L117}

\bibitem[{{Svensson} \& {Zdziarski}(1994)}]{Svensson1994}
{Svensson}, R., \& {Zdziarski}, A.~A. 1994, \apj, 436, 599,
  \dodoi{10.1086/174934}

\bibitem[{{Taverna} {et~al.}(2020){Taverna}, {Zhang}, {Dov{\v{c}}iak},
  {Bianchi}, {Bursa}, {Karas}, \& {Matt}}]{Taverna2020}
{Taverna}, R., {Zhang}, W., {Dov{\v{c}}iak}, M., {et~al.} 2020, \mnras, 493,
  4960, \dodoi{10.1093/mnras/staa598}

\bibitem[{{Tomsick} {et~al.}(2018){Tomsick}, {Parker}, {Garc{\'\i}a},
  {Yamaoka}, {Barret}, {Chiu}, {Clavel}, {Fabian}, {F{\"u}rst}, {Gandhi},
  {Grinberg}, {Miller}, {Pottschmidt}, \& {Walton}}]{Tomsick2018}
{Tomsick}, J.~A., {Parker}, M.~L., {Garc{\'\i}a}, J.~A., {et~al.} 2018, \apj,
  855, 3, \dodoi{10.3847/1538-4357/aaaab1}

\bibitem[{{Verner} {et~al.}(1996){Verner}, {Ferland}, {Korista}, \&
  {Yakovlev}}]{Verner1996}
{Verner}, D.~A., {Ferland}, G.~J., {Korista}, K.~T., \& {Yakovlev}, D.~G. 1996,
  \apj, 465, 487, \dodoi{10.1086/177435}

\bibitem[{{Wilms} {et~al.}(2000){Wilms}, {Allen}, \& {McCray}}]{Wilms2000}
{Wilms}, J., {Allen}, A., \& {McCray}, R. 2000, \apj, 542, 914,
  \dodoi{10.1086/317016}

\bibitem[{{Yang} {et~al.}(2010){Yang}, {Brocksopp}, {Corbel}, {Paragi},
  {Tzioumis}, \& {Fender}}]{Yang2010}
{Yang}, J., {Brocksopp}, C., {Corbel}, S., {et~al.} 2010, \mnras, 409, L64,
  \dodoi{10.1111/j.1745-3933.2010.00948.x}

\bibitem[{{Yang} {et~al.}(2011){Yang}, {Paragi}, {Corbel}, {Gurvits},
  {Campbell}, \& {Brocksopp}}]{Yang2011}
{Yang}, J., {Paragi}, Z., {Corbel}, S., {et~al.} 2011, \mnras, 418, L25,
  \dodoi{10.1111/j.1745-3933.2011.01136.x}

\bibitem[{{Young} {et~al.}(1998){Young}, {Ross}, \& {Fabian}}]{Young1998}
{Young}, A.~J., {Ross}, R.~R., \& {Fabian}, A.~C. 1998, \mnras, 300, L11,
  \dodoi{10.1046/j.1365-8711.1998.02058.x}

\bibitem[{{Zdziarski} \& {De Marco}(2020)}]{Zdziarski2020a}
{Zdziarski}, A.~A., \& {De Marco}, B. 2020, \apjl, 896, L36,
  \dodoi{10.3847/2041-8213/ab9899}

\bibitem[{{Zdziarski} {et~al.}(2021{\natexlab{a}}){Zdziarski}, {De Marco},
  {Szanecki}, {Nied{\'z}wiecki}, \& {Markowitz}}]{Zdziarski2021a}
{Zdziarski}, A.~A., {De Marco}, B., {Szanecki}, M., {Nied{\'z}wiecki}, A., \&
  {Markowitz}, A. 2021{\natexlab{a}}, \apj, 906, 69,
  \dodoi{10.3847/1538-4357/abca9c}

\bibitem[{{Zdziarski} {et~al.}(2021{\natexlab{b}}){Zdziarski}, {Dzie{\l}ak},
  {De Marco}, {Szanecki}, \& {Nied{\'z}wiecki}}]{Zdziarski2021b}
{Zdziarski}, A.~A., {Dzie{\l}ak}, M.~A., {De Marco}, B., {Szanecki}, M., \&
  {Nied{\'z}wiecki}, A. 2021{\natexlab{b}}, \apjl, 909, L9,
  \dodoi{10.3847/2041-8213/abe7ef}

\bibitem[{{Zdziarski} {et~al.}(1996){Zdziarski}, {Johnson}, \&
  {Magdziarz}}]{Zdziarski1996}
{Zdziarski}, A.~A., {Johnson}, W.~N., \& {Magdziarz}, P. 1996, \mnras, 283,
  193, \dodoi{10.1093/mnras/283.1.193}

\bibitem[{{Zdziarski} {et~al.}(2020){Zdziarski}, {Szanecki}, {Poutanen},
  {Gierli{\'n}ski}, \& {Biernacki}}]{Zdziarski2020c}
{Zdziarski}, A.~A., {Szanecki}, M., {Poutanen}, J., {Gierli{\'n}ski}, M., \&
  {Biernacki}, P. 2020, \mnras, 492, 5234, \dodoi{10.1093/mnras/staa159}

\bibitem[{{Zimmerman} {et~al.}(2005){Zimmerman}, {Narayan}, {McClintock}, \&
  {Miller}}]{Zimmerman2005}
{Zimmerman}, E.~R., {Narayan}, R., {McClintock}, J.~E., \& {Miller}, J.~M.
  2005, \apj, 618, 832, \dodoi{10.1086/426071}

\bibitem[{{{\.Z}ycki} {et~al.}(1999){{\.Z}ycki}, {Done}, \&
  {Smith}}]{Zycki1999}
{{\.Z}ycki}, P.~T., {Done}, C., \& {Smith}, D.~A. 1999, \mnras, 309, 561,
  \dodoi{10.1046/j.1365-8711.1999.02885.x}

\end{thebibliography}
%
%
%
%

\end{document}